\documentclass[12pt]{article}

\usepackage[english]{babel}
\usepackage[utf8x]{inputenc}
\usepackage[T1]{fontenc}

\usepackage[a4paper,top=3cm,bottom=2cm,left=2.5cm,right=2.5cm,marginparwidth=1.75cm]{geometry}

\usepackage{graphicx}
\usepackage{float}
\usepackage[colorlinks=true, allcolors=blue]{hyperref}
\usepackage{setspace}
\usepackage{subfigure}
\usepackage{tcolorbox}

\title{Open Source Development Around the World: A Comparative Study}
\author{Thaís Mombach$^1$, Marco Tulio Valente$^1$, Cuiting Chen$^2$, \\ Magiel Bruntink$^2$, Gustavo Pinto$^3$}

\date{{\normalsize $^1$Federal University of Minas Gerais, Brazil; $^2$Software Improvement Group, Netherlands; $^3$Federal University of Par\'a, Brazil} \\
{\small{\{mtov,thaismombach\}@dcc.ufmg.br}, \{m.bruntink,c.chen\}@sig.eu, gpinto@ufpa.br }}

\begin{document}
\maketitle

\begin{abstract}
\noindent Open source software has an increasing importance in our modern society, providing basic services to other software systems and also supporting the rapid development of a variety of end-user applications. 
Recently, world-wide code sharing platforms, like GitHub, are also contributing to open source's growth. However, little is known on how this growth is distributed around the world and about the characteristics of the projects developed in different countries. In this article, we provide a characterization of 2,648 open source projects developed in 20 countries. We reveal the number of projects per country, the popularity and programming language of each country's project and also show how the number of projects in a country correlates to its GDP. Finally, we assess the maintainability and internal code quality of the studied projects, using a tool called BetterCodeHub. 
\end{abstract}

\section{Introduction}


Almost 10 years ago, Bjarne Stroustrup proclaimed that ``our civilization runs on software''~\cite{stroustrup2014programming}.
Today, we can rephrase his declaration, stating that our civilization runs on  {\em open source} software. For example, a variety of popular and critical
applications are currently available under open source licenses, including operating systems, 
compilers, databases, and web servers. Furthermore, 
end-user applications are rapidly designed and released due to the reuse of code provided by open source 
libraries and frameworks.
As a consequence, the costs of
running software intensive organizations dropped dramatically in the last years. For example, the costs of launching a startup declined 90\% in 
a period of 20 years, since infrastructure
software nowadays is nearly free~\cite{nadia2016roads}. 

Another key characteristic of open source projects is their global nature and reach. For example, the Linux kernel includes contributions of more than 15K developers, from at least 1,500 companies and 71 countries~\cite{dempsey2002open,linux2017}. Recently, modern source code sharing platforms, like GitHub, allow geographically dispersed developers to launch, maintain, and deploy open source projects. GitHub standardizes a workflow to handle contributions from external developers, besides providing services to manage maintenance issues, to document projects, to support code reviews and other quality assurance tasks. 


However, it is still unclear whether open source project's characteristics such as popularity and maintainability depend on the country on which the project was built. 
In this paper, we investigate 2,648 popular GitHub projects, maintained by individuals or organizations from 20 countries. For each country, we reveal the number of projects and the main programming language used by its projects, we correlate the number of projects with a country's GDP, and we study projects' popularity in terms of the number of stars. We also investigate maintainability metrics, including Separation of Concerns, Clean Code, and Automated Tests. Our ultimate goal is to provide a picture on the state of open source development around the world. This picture can help governments, entrepreneurs, and non-profit organizations to better develop strategies that foster and support open source projects in their countries. 

\section{Study Design}

We focus on popular projects, with a large base of users and developers. 
Therefore, as of November 2017, we selected the top-10K open source projects on GitHub by the number of stars. GitHub stars are similar to {\em likes} in other social networks and they are considered  a reliable proxy for the popularity of GitHub projects~\cite{popularity2016}.  

GitHub does not directly provide the geographic location of a project. However, location is a meta-data of GitHub accounts. For example, suppose the project {\sc aserg-ufmg/jscity}. The owner of this project is {\sc aserg-ufmg}, which is an organizational account on GitHub. In {\sc aserg-ufmg}'s profile, it is informed that this organization is located in {\em Belo Horizonte, Brazil}.\footnote{GitHub projects have a unique owner; but contributors can be of different countries, as studied in Section~\ref{sec:international-domestic}.}
Therefore, we consider {\sc aserg-ufmg/jscity} as a Brazilian project. However, this approach has some limitations; for example, 3,311 projects (33\%) are owned by accounts with an empty location field. We removed these projects, resulting in a list of 6,689 projects.
We also removed 645 projects that are not software systems (e.g.,~projects referring to programming books, tutorials, awesome lists, etc). To remove these projects, we first computed their number of lines of code, using a tool called {\tt cloc} ({\url{https://github.com/AlDanial/cloc}). We configured this tool to consider source code written in the top-100 most popular programming languages, according to the TIOBE index (\url{https://www.tiobe.com/tiobe-index}). We then removed projects that have no lines of code, as returned by {\tt cloc}.

For the remaining 6,044 projects, we implemented and used a script that attempts to match a project location to a list of country names in English; we were successful for 2,870 projects (47\%). Finally, the first author of this paper inspected the location of the remaining projects, aiming to manually associate them to countries, which was possible in the case of other 2,518 projects (42\%).
For example, the location of one project just mentioned {\em Broadcasting House, London}, which is the BBC headquarters in London; therefore, its country was manually defined as UK. 
The remaining 656 projects (11\%) include locations which are not countries (e.g.,~{\em The Earth}) or locations mentioning more than one country (e.g.,~{\em Canada \& France}).



After following these steps, we were able to identify the country of 5,388 projects, which are distributed over 77 countries (see Figure~\ref{fig:map_countries}). United States has the largest number of projects (2,302 projects, 42\%), which is three times greater than the second country (China). However, we decided to remove the United States from our analysis, because the computation of the maintainability metrics---as detailed in Section~\ref{sec:bch}---requires some manual configuration steps. Therefore, we decided to compute these metrics for the next 20 countries, to maximize the number  and geographical distribution of the analyzed countries. As a result, we study 2,648 open source projects. These projects are presented in Figure~\ref{fig:ini_countries}. 
The number of projects in these countries ranges from 754 (China) to 32 (Norway).

\begin{figure}[!ht]
\centering
\includegraphics[width=\textwidth]{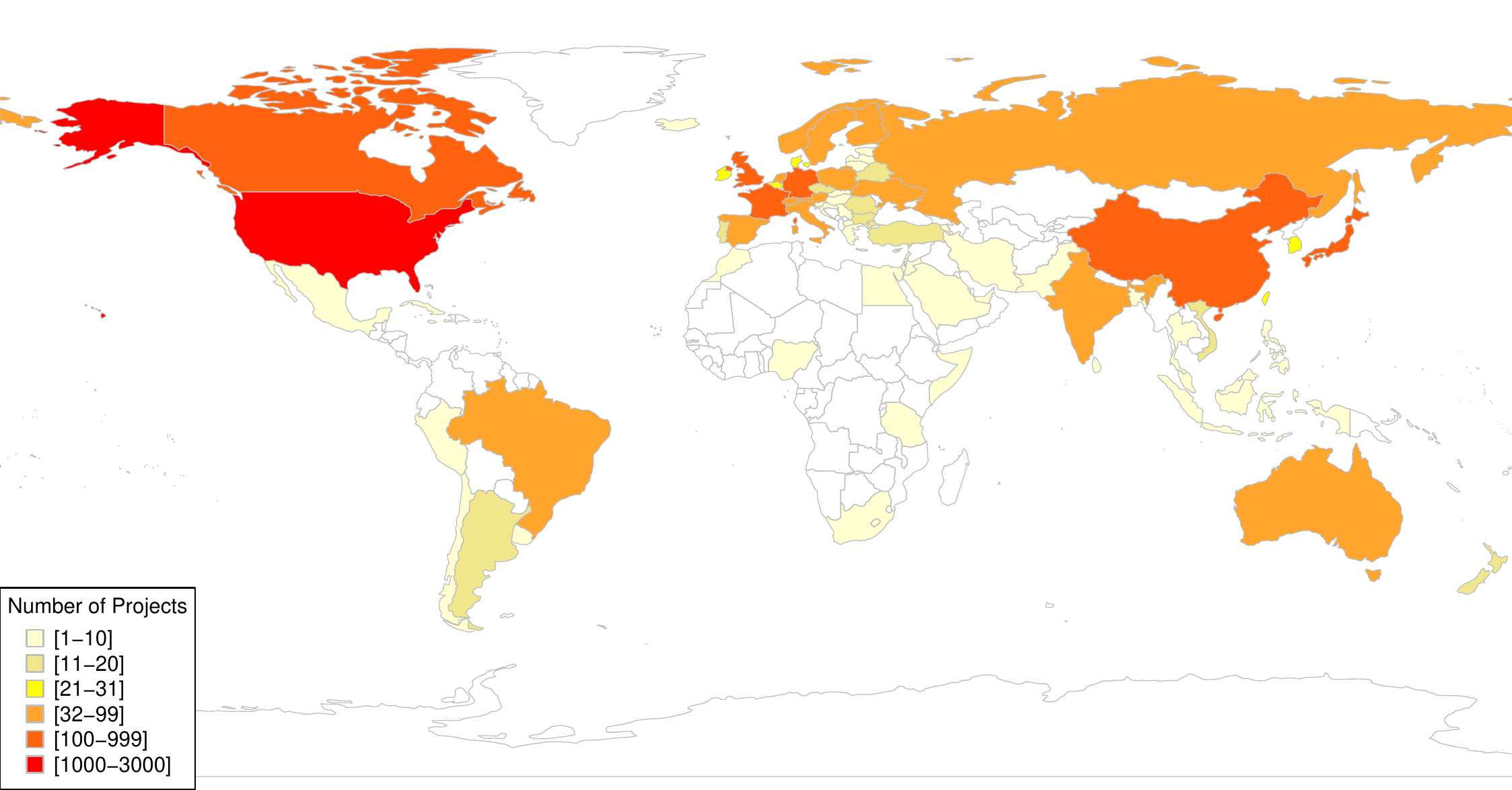}
\caption{Number of GitHub projects per country}
\label{fig:map_countries}
\end{figure}

\begin{figure}[!ht]
\centering
\includegraphics[width=\textwidth]{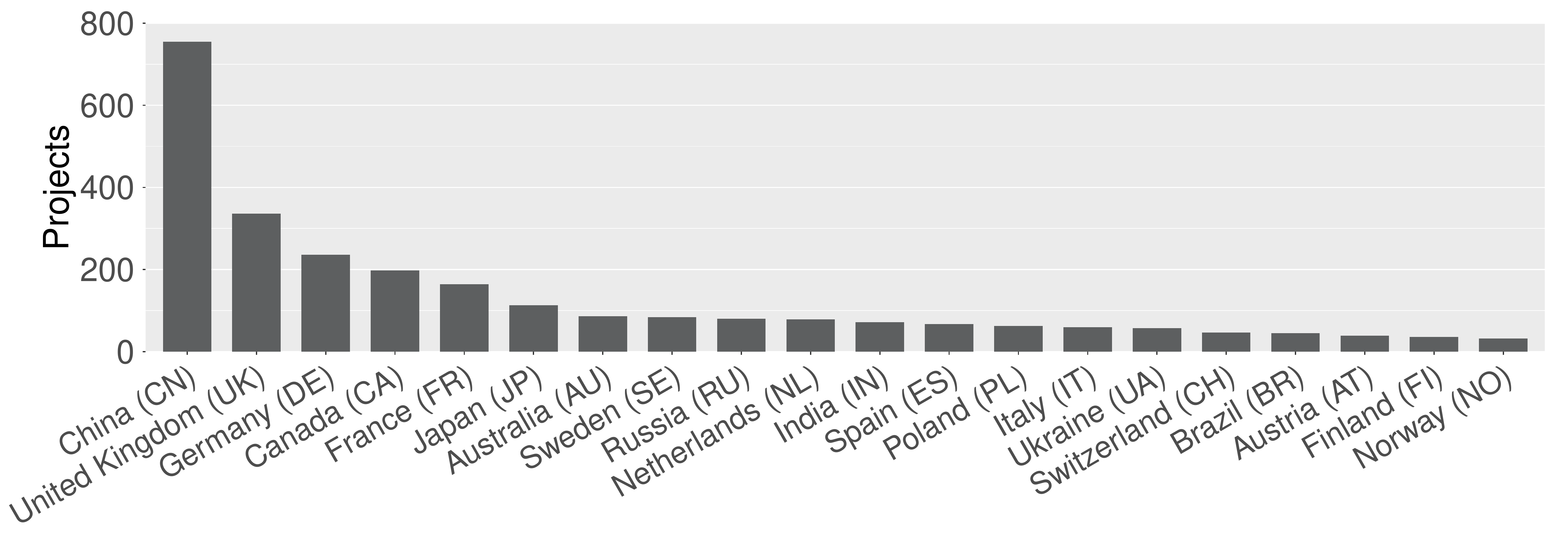}
\caption{Countries studied in this work}
\label{fig:ini_countries}
\end{figure}

\section{Results}

\subsection{Correlation with GDP}

Figure \ref{fig:gdp_relation} correlates 
the number of projects and the Growth Domestic Product (GDP) of the studied countries. The GDP values refer to the most recent year available for each country, as estimated by the International Monetary Fund (\url{http://www.imf.org/external/index.htm}). We excluded China from this figure, to ease visualization (China has the largest GDP among the studied projects and also the largest number of projects). As we can observe in the figure,
there is a strong and positive monotonic correlation among these two variables (Spearman's $\rho= 0.75$, {\em p-value} $\leq 0.01$). 
The four countries with the highest ratio of projects per GDP (in billions) are Ukraine (0.61), Sweden (0.16), and Finland (0.15). The countries with the lowest ratio are Japan (0.02), India (0.03), Italy (0.03), and Brazil (0.03). Therefore, these latter countries can define measures and policies to increase the visibility of their open source projects, e.g.,~governments, universities, and open source foundations can organize users meetings, hackathons and similar events, to better promote open source practices among local developers.



\begin{figure}[!ht]
\centering
\includegraphics[width=0.7\textwidth]{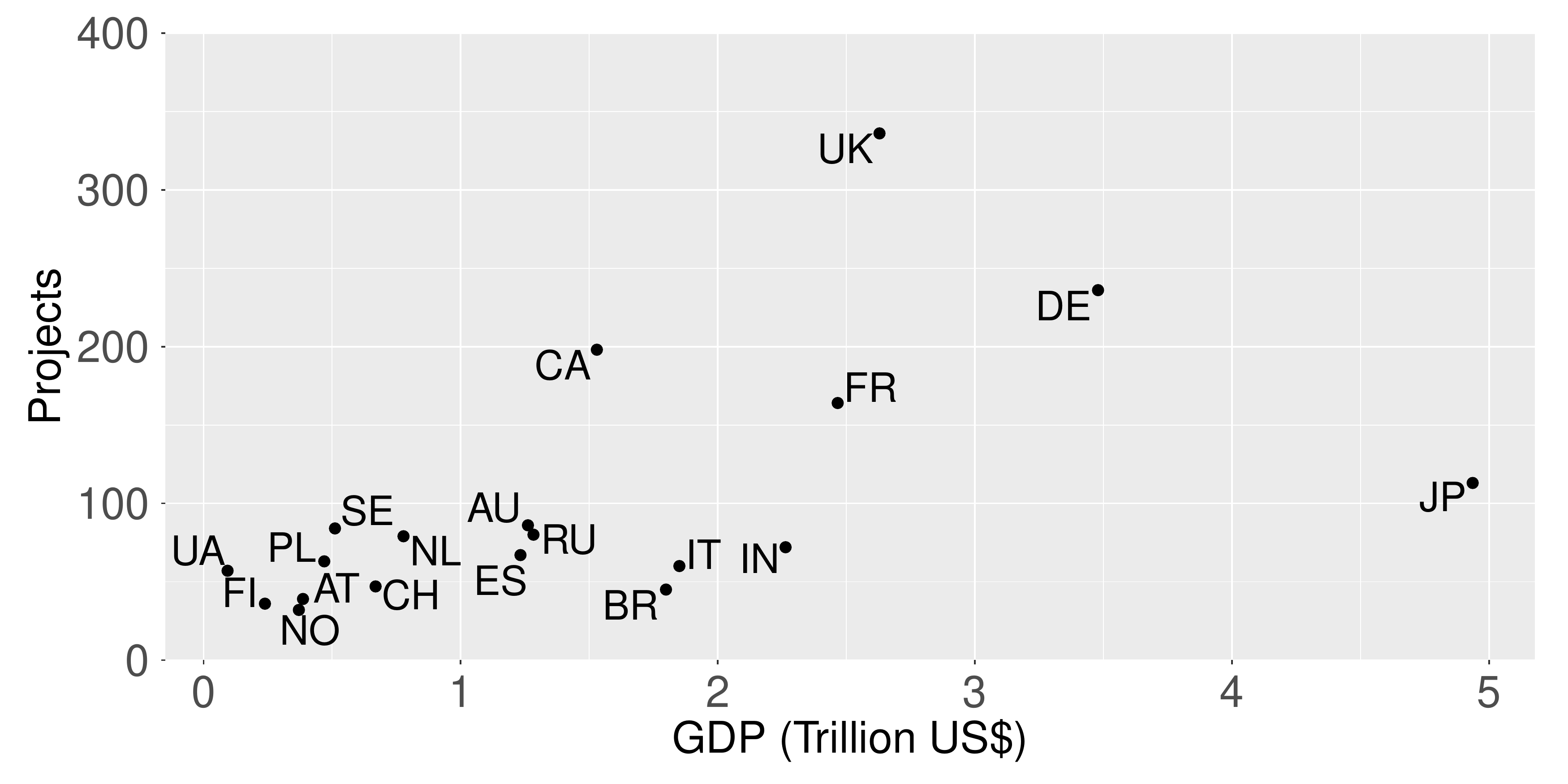}
\caption{GDP vs number of projects (excluding China, to ease visualization)}
\label{fig:gdp_relation}
\end{figure}

\subsection{Popularity}

Figure~\ref{fig:popularity_percountry} presents the distribution of the number of stars for the projects in each studied country. If we focus on the median values for each country, there are no major differences. The median values ranges from 1,705.5 stars (Switzerland) to 2,406.5 stars (Norway). 
This result suggests that in all countries there are developers who managed to create popular projects; therefore, by means of planned actions and policies it might be possible to attract other developers to open source, particularly in countries with a large population and a relatively small number of popular projects (e.g., India and Brazil).


\begin{figure}[ht]
\centering
\includegraphics[width=\textwidth]{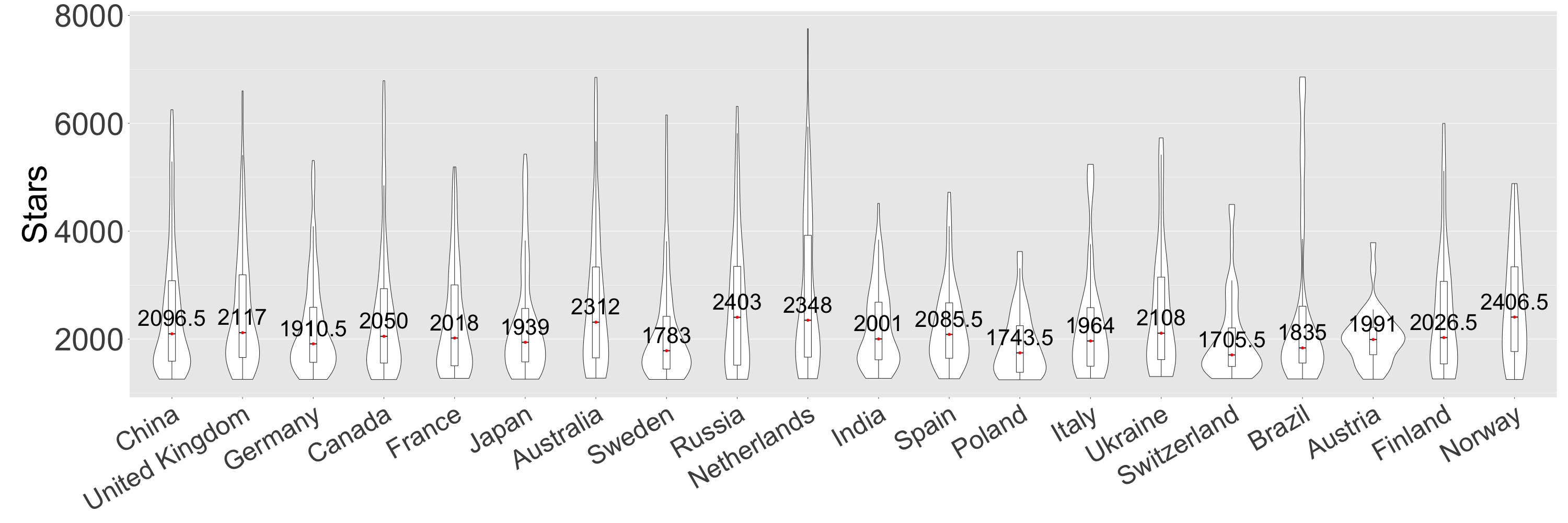}
\caption{Popularity in terms of the number of stars
}
\label{fig:popularity_percountry}
\end{figure}

\subsection{Programming Languages}

Figure~\ref{fig:language} shows 
the top-3 most popular programming languages for each country, i.e., for each country we computed the primary programming language of its projects. JavaScript is the most popular language in 18 countries and Java is the most popular one in the remaining two countries (China and Ukraine). Java is also ranked among the top-3 languages of 16 countries; the exceptions are Canada, Italy, Switzerland, and Brazil.
This result reveals the skills and preferences of the developers in each country, since programming languages are often used to implement particular kinds of system. For example, Apple software is usually implemented in Objective-C (which is a popular programming language in  China, United Kingdom, Switzerland, and Austria) or Swift (which is popular in Japan, Sweden, Poland, Italy, Ukraine, and Norway). In other words, countries tend to present an installed competence in particular programming languages and systems; probably, these are the most recommended technologies to receive attention by governments and open source organizations, in order to foster open source practices in their countries.


\begin{figure}[!ht]
\centering
\subfigure[CN]{
\includegraphics[width=0.17\textwidth]{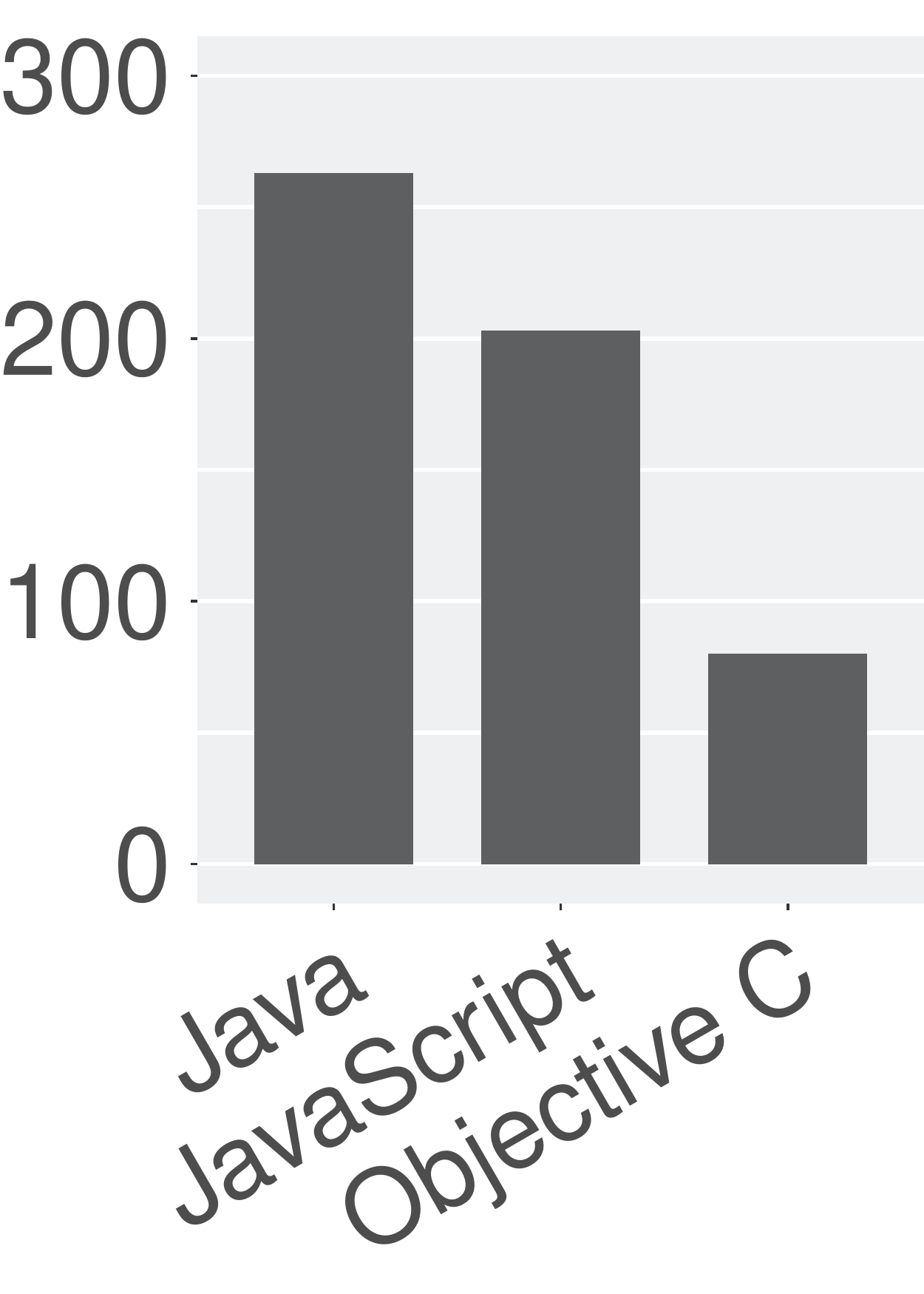}
\label{subfig:language_china}
}
\subfigure[UK]{
\includegraphics[width=0.17\textwidth]{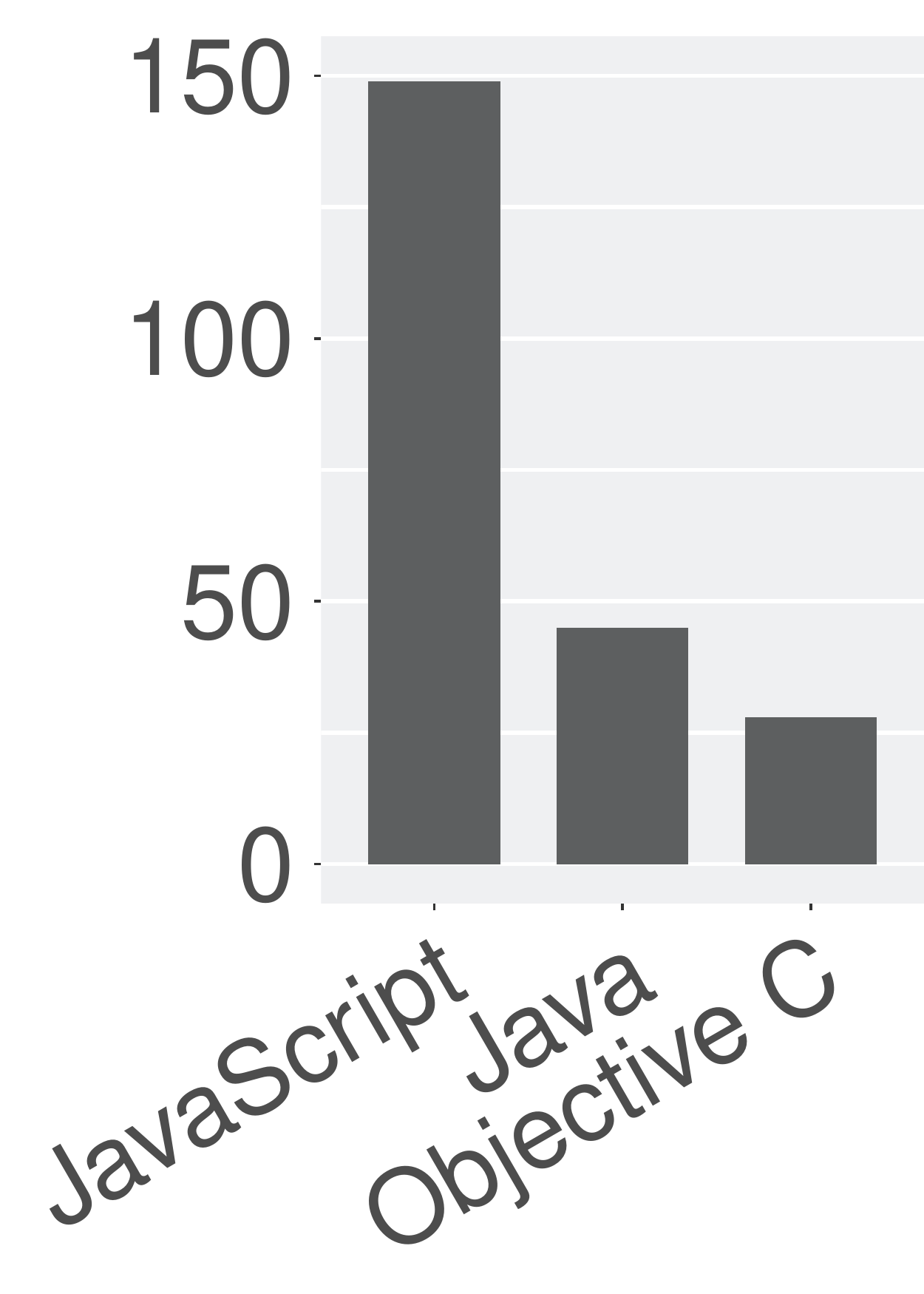}
\label{subfig:language_unitedkingdom}
}
\subfigure[DE]{
\includegraphics[width=0.17\textwidth]{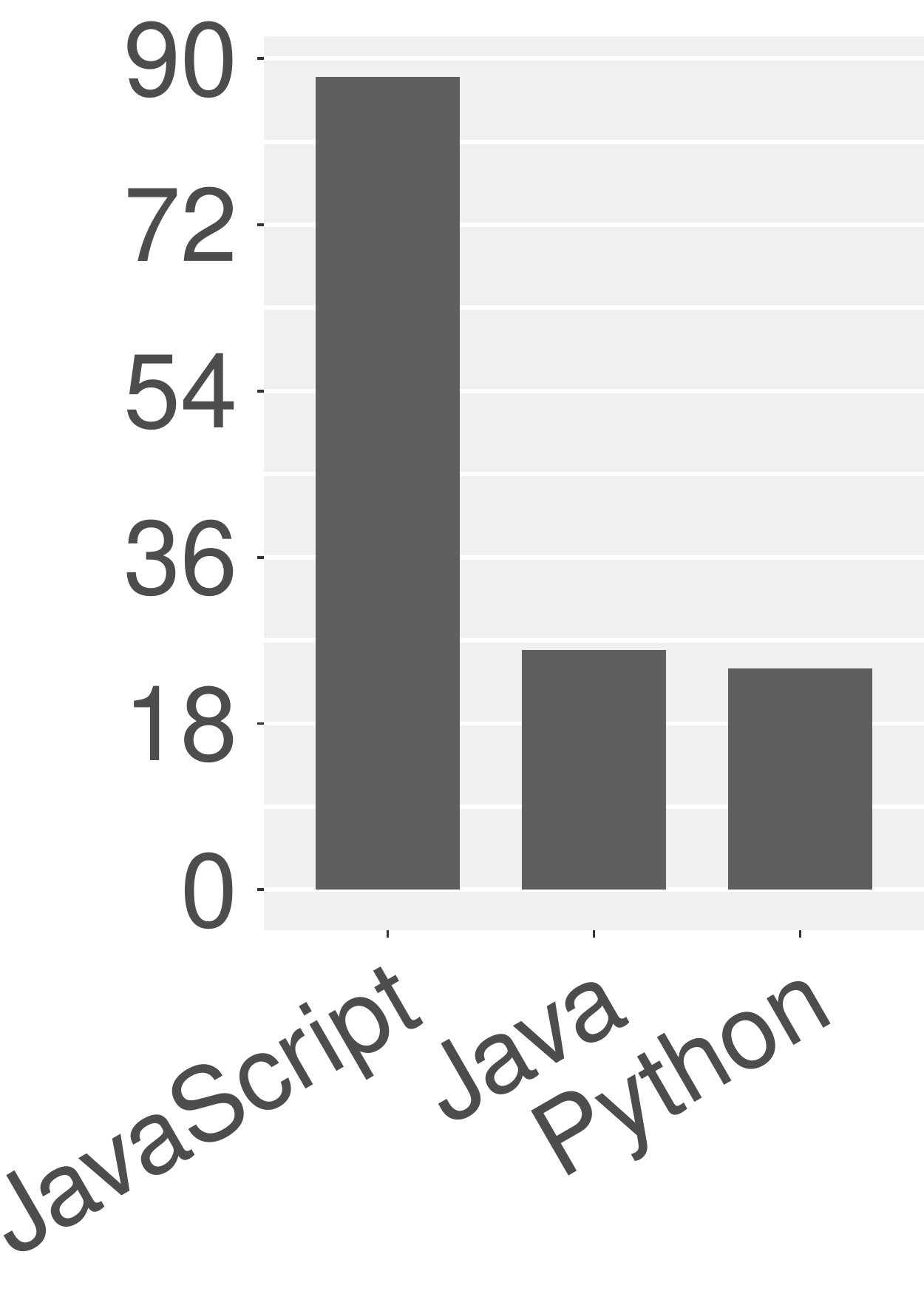}
\label{subfig:language_germany}
}
\subfigure[CA]{
\includegraphics[width=0.17\textwidth]{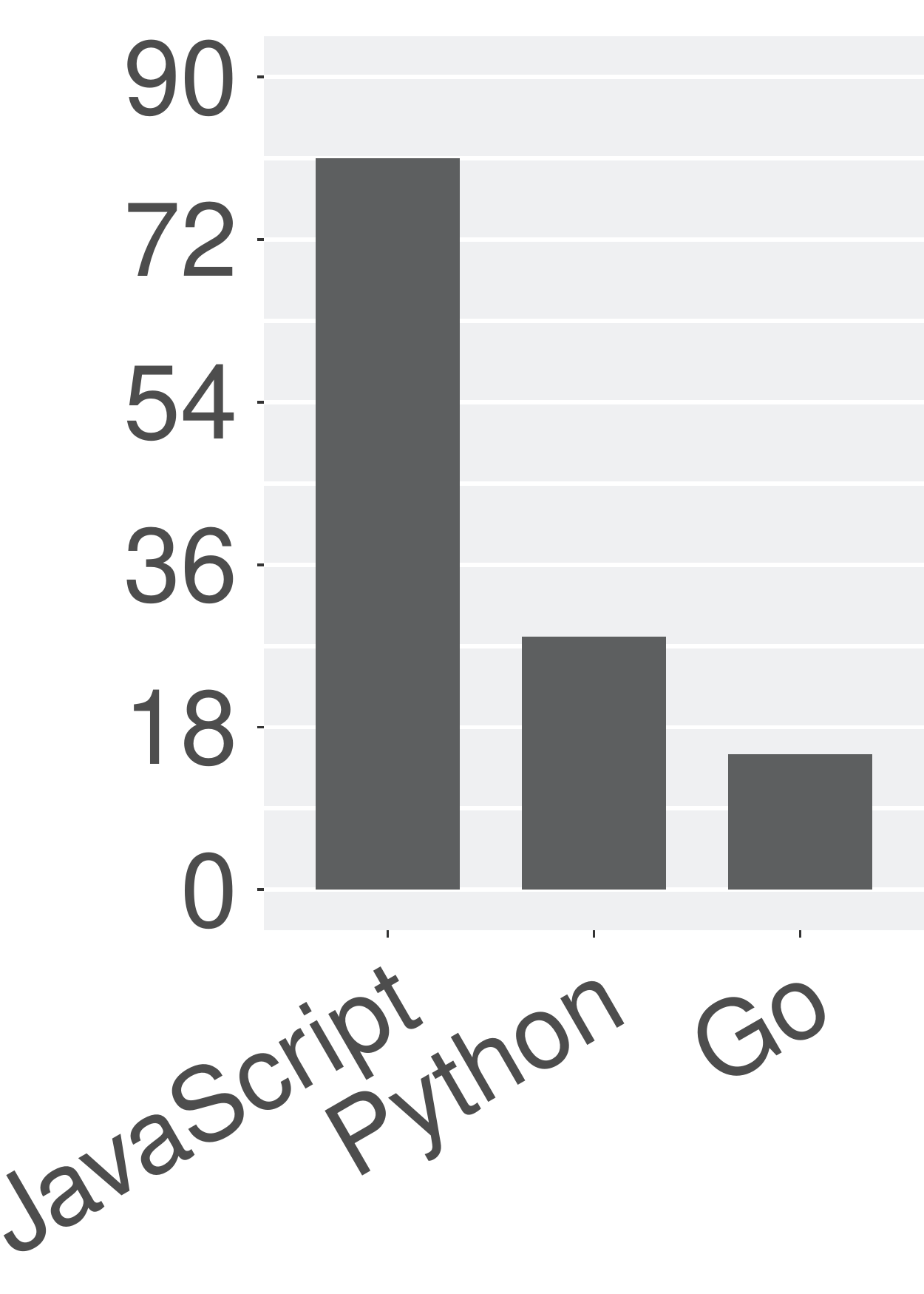}
\label{subfig:language_canada}
}
\subfigure[FR]{
\includegraphics[width=0.17\textwidth]{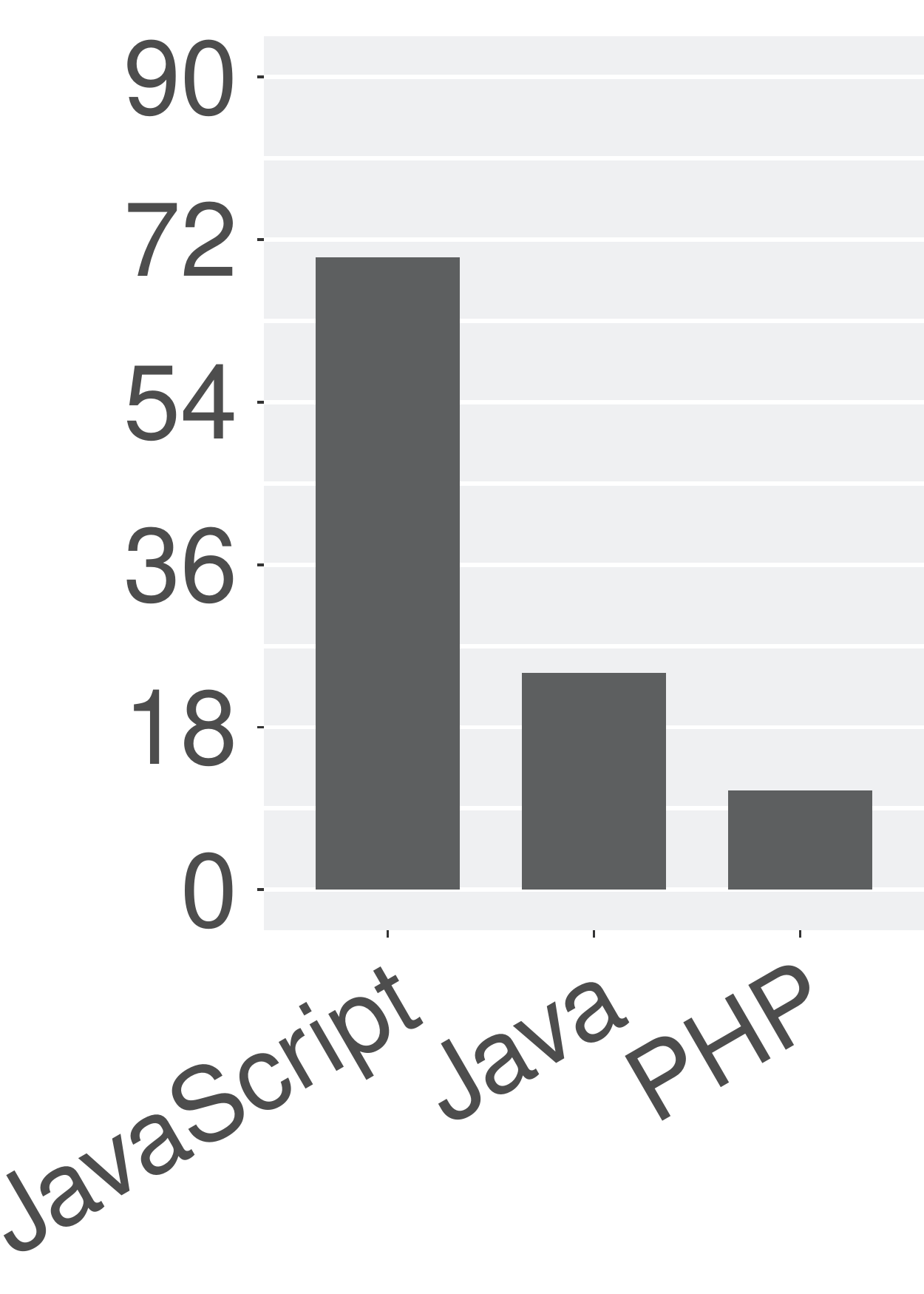}
\label{subfig:language_france}
}
\subfigure[JP]{
\includegraphics[width=0.17\textwidth]{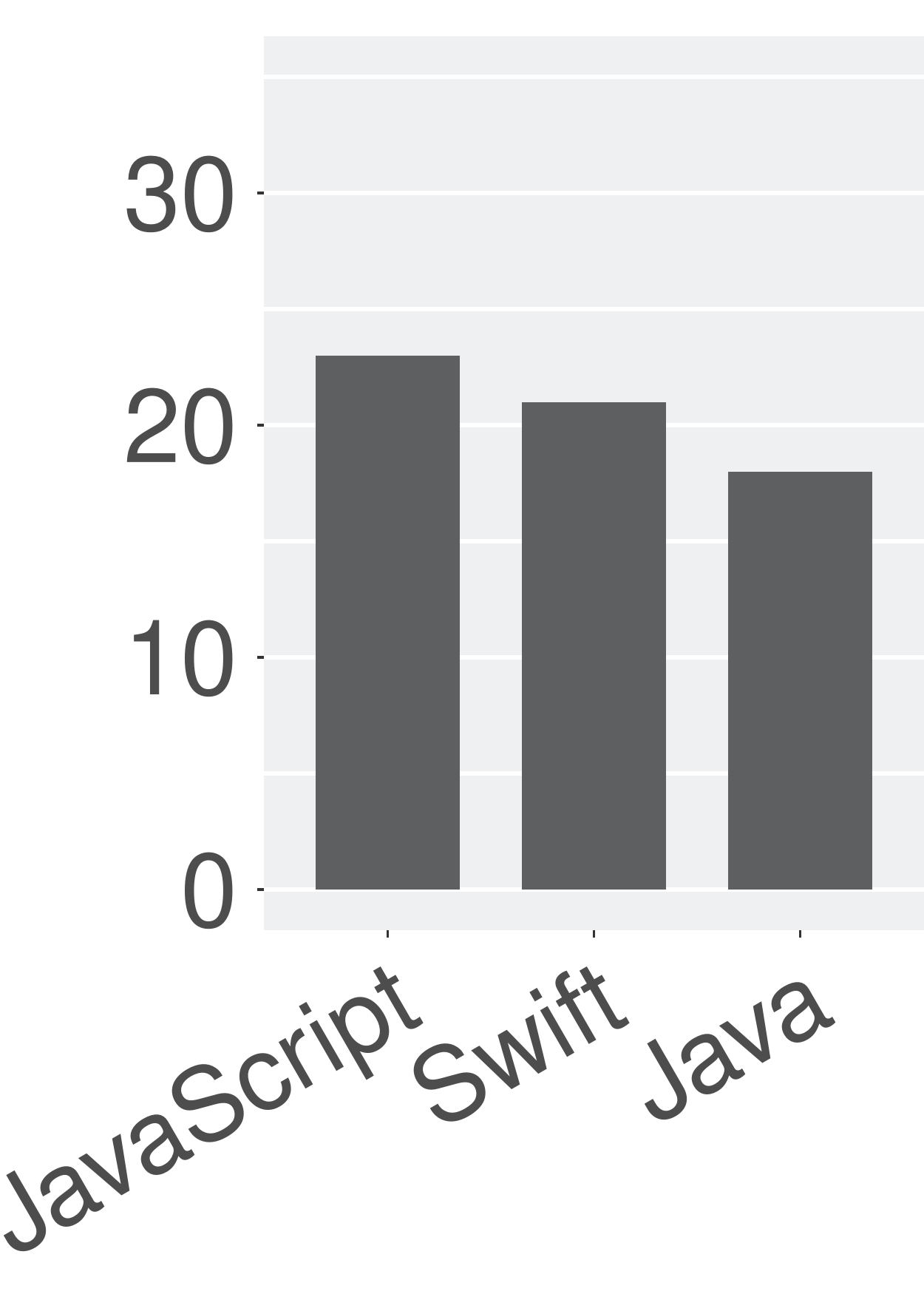}
\label{subfig:language_japan}
}
\subfigure[AU]{
\includegraphics[width=0.17\textwidth]{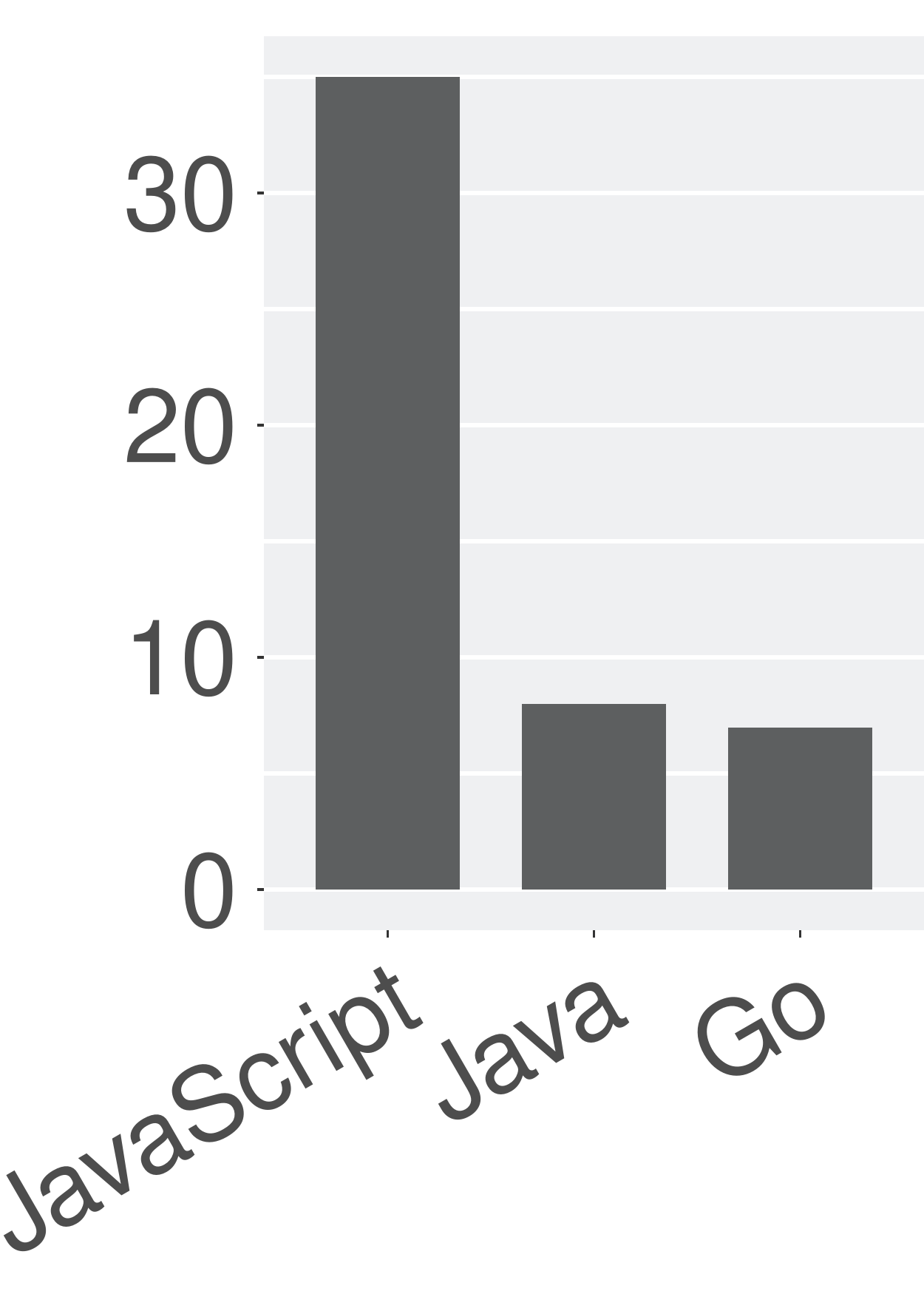}
\label{subfig:language_australia}
}
\subfigure[SE]{
\includegraphics[width=0.17\textwidth]{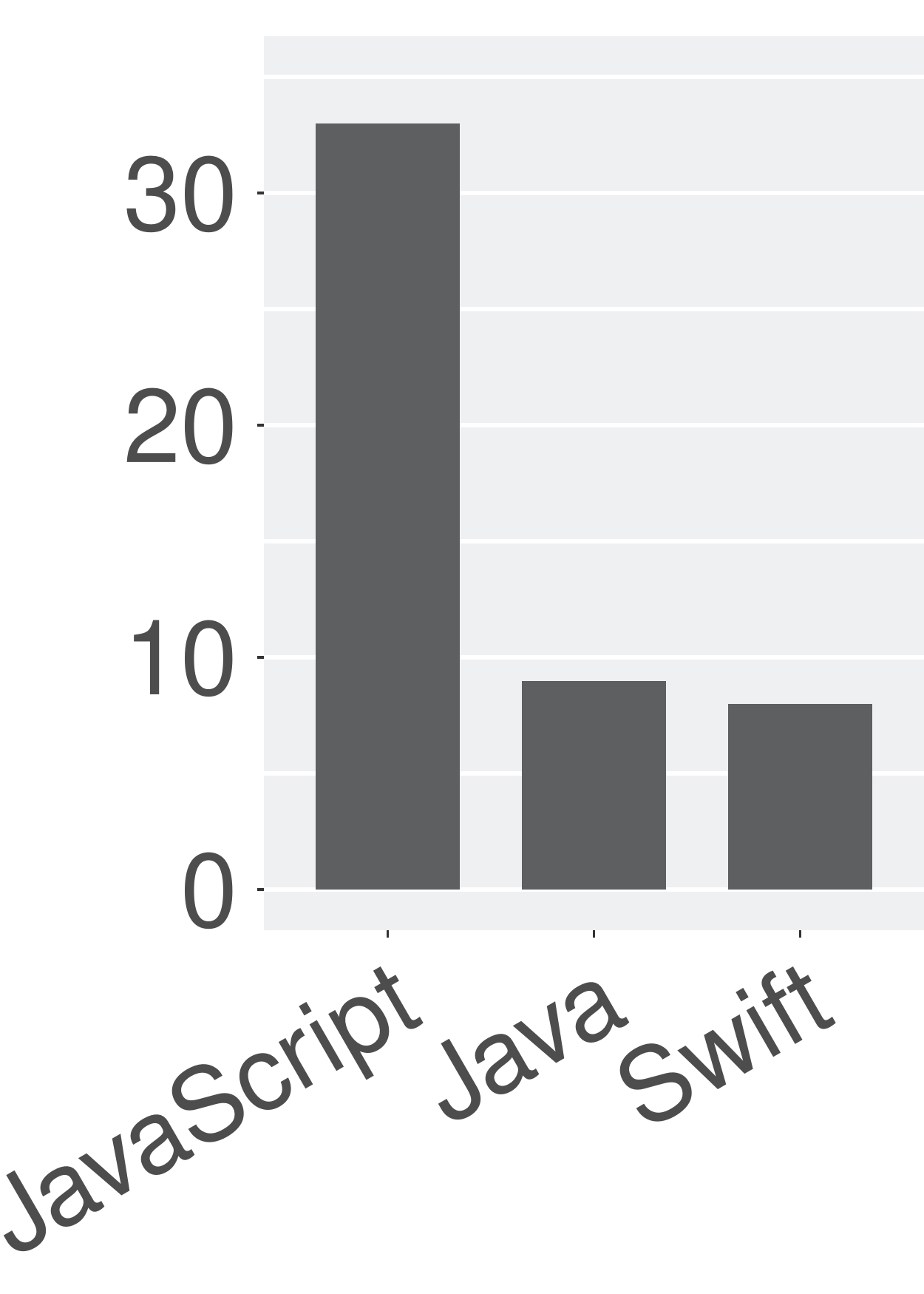}
\label{subfig:language_sweden}
}
\subfigure[NL]{
\includegraphics[width=0.17\textwidth]{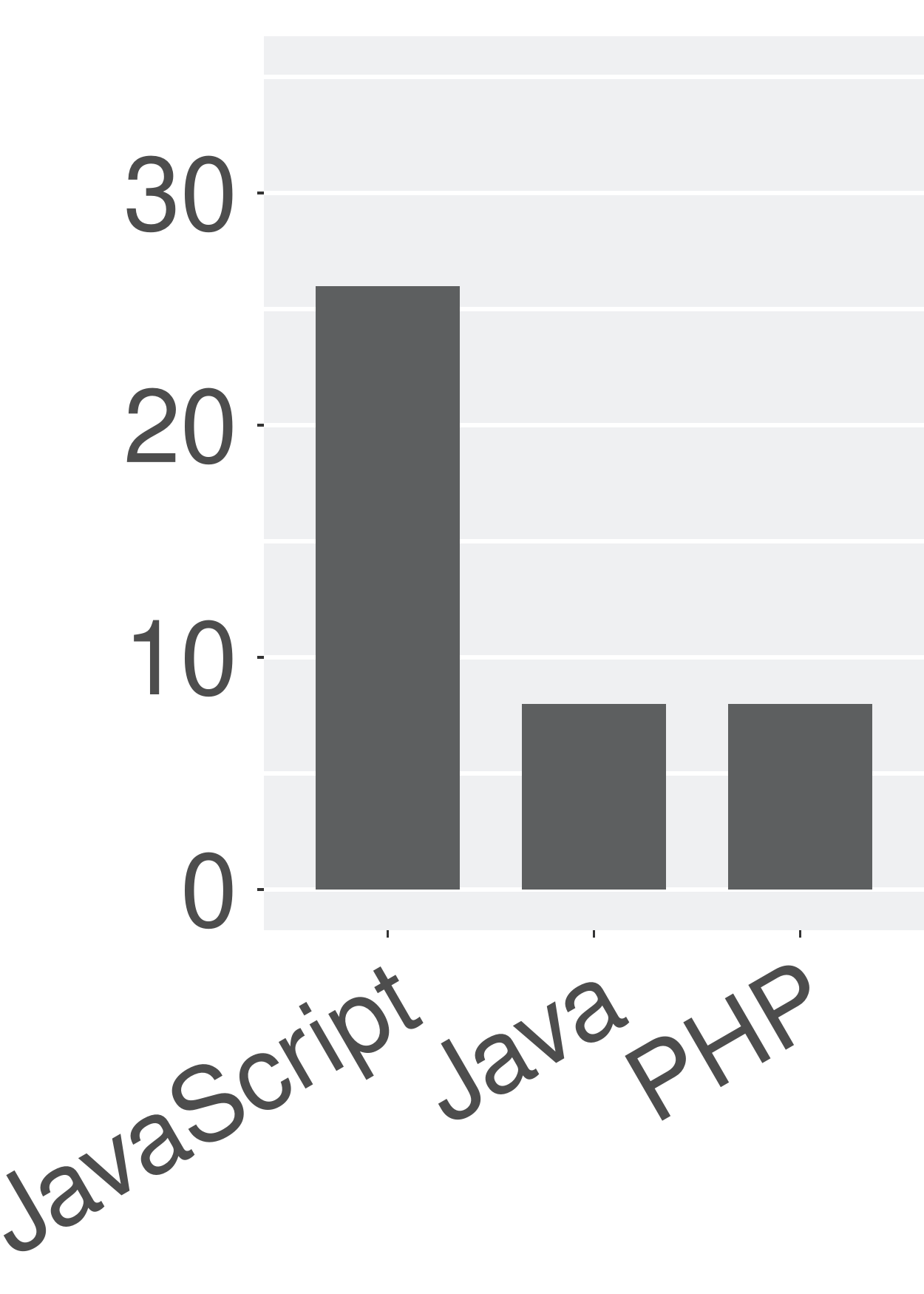}
\label{subfig:language_netherlands}
}
\subfigure[RU]{
\includegraphics[width=0.17\textwidth]{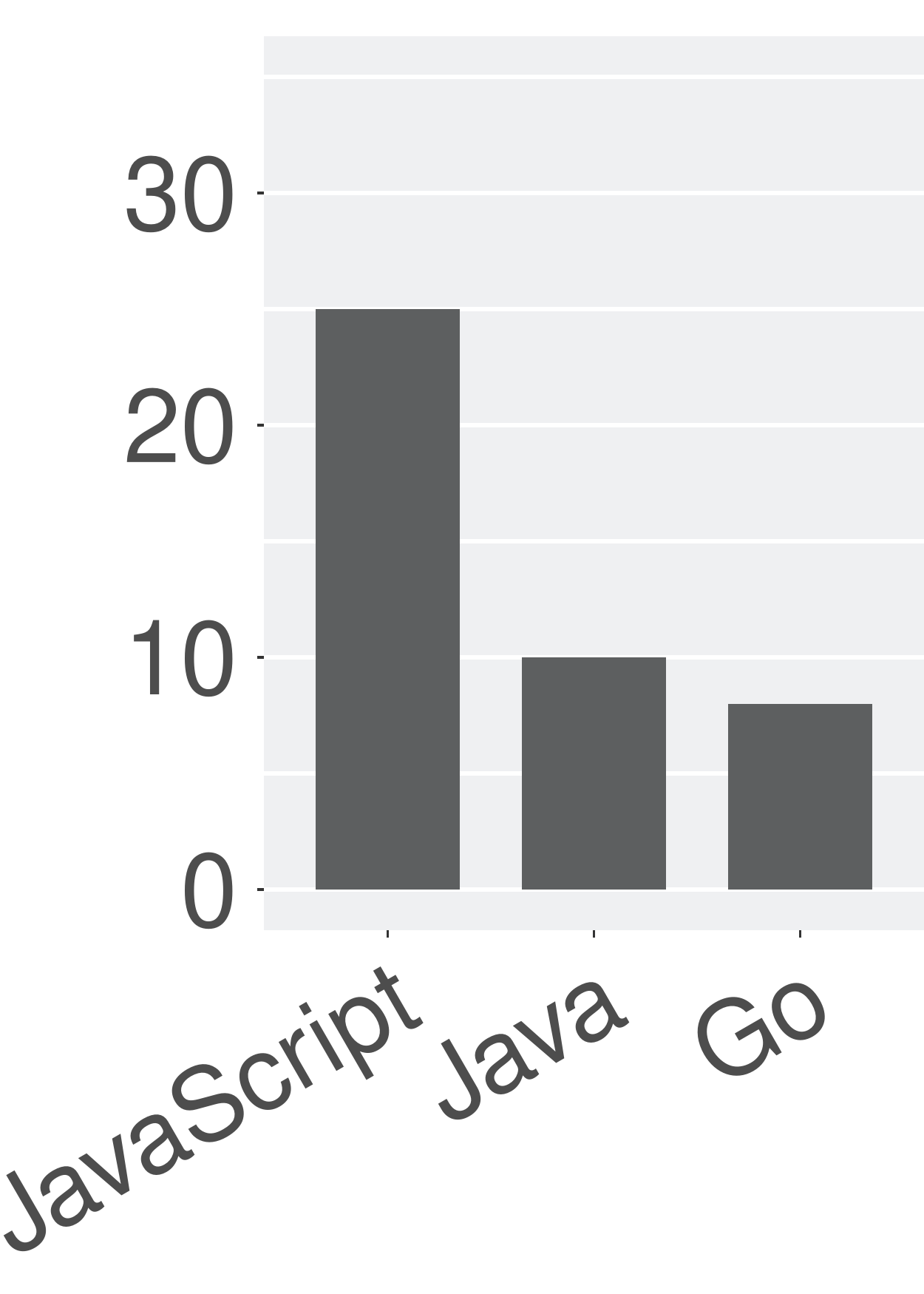}
\label{subfig:language_russia}
}
\subfigure[IN]{
\includegraphics[width=0.17\textwidth]{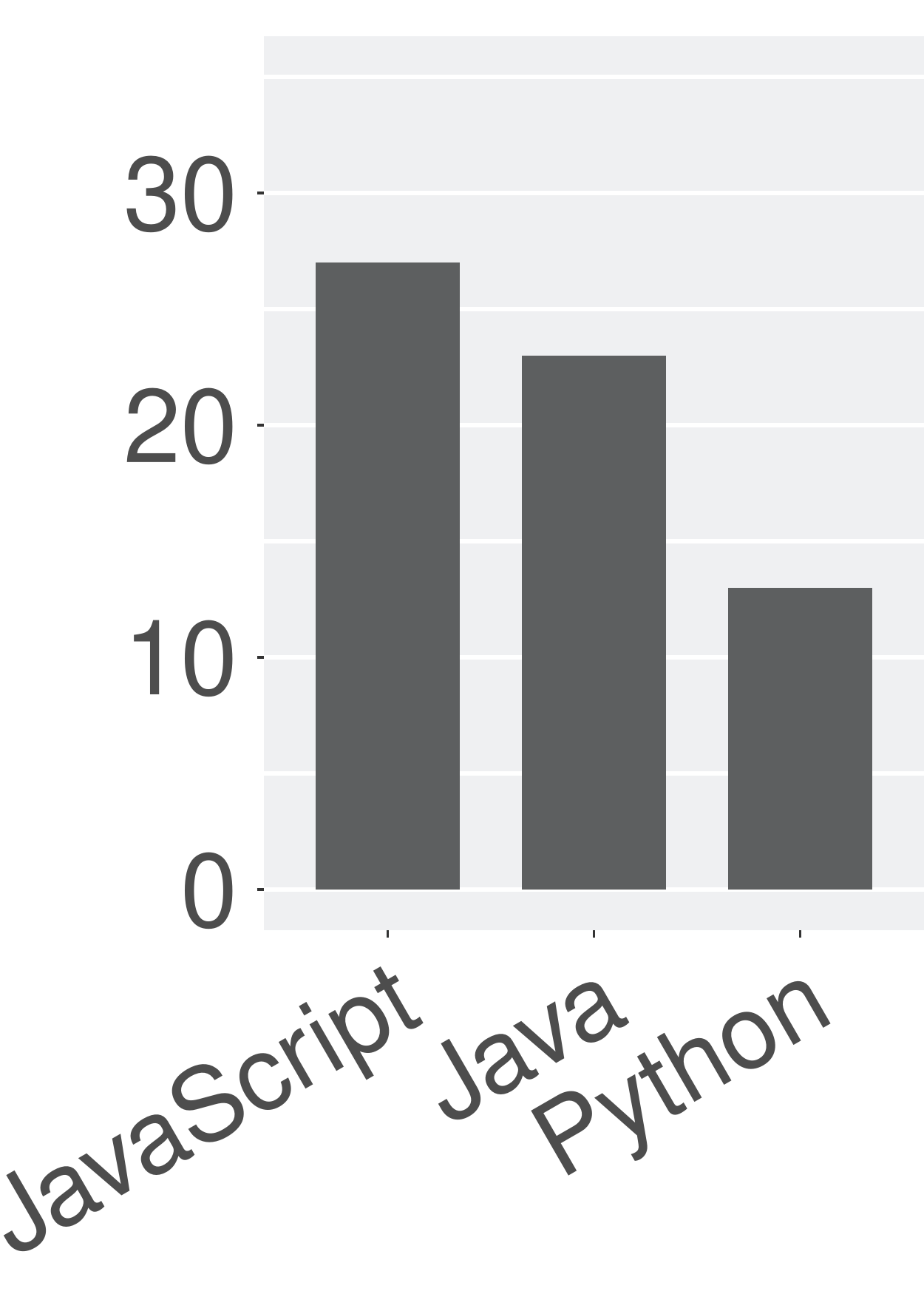}
\label{subfig:language_india}
}
\subfigure[ES]{
\includegraphics[width=0.17\textwidth]{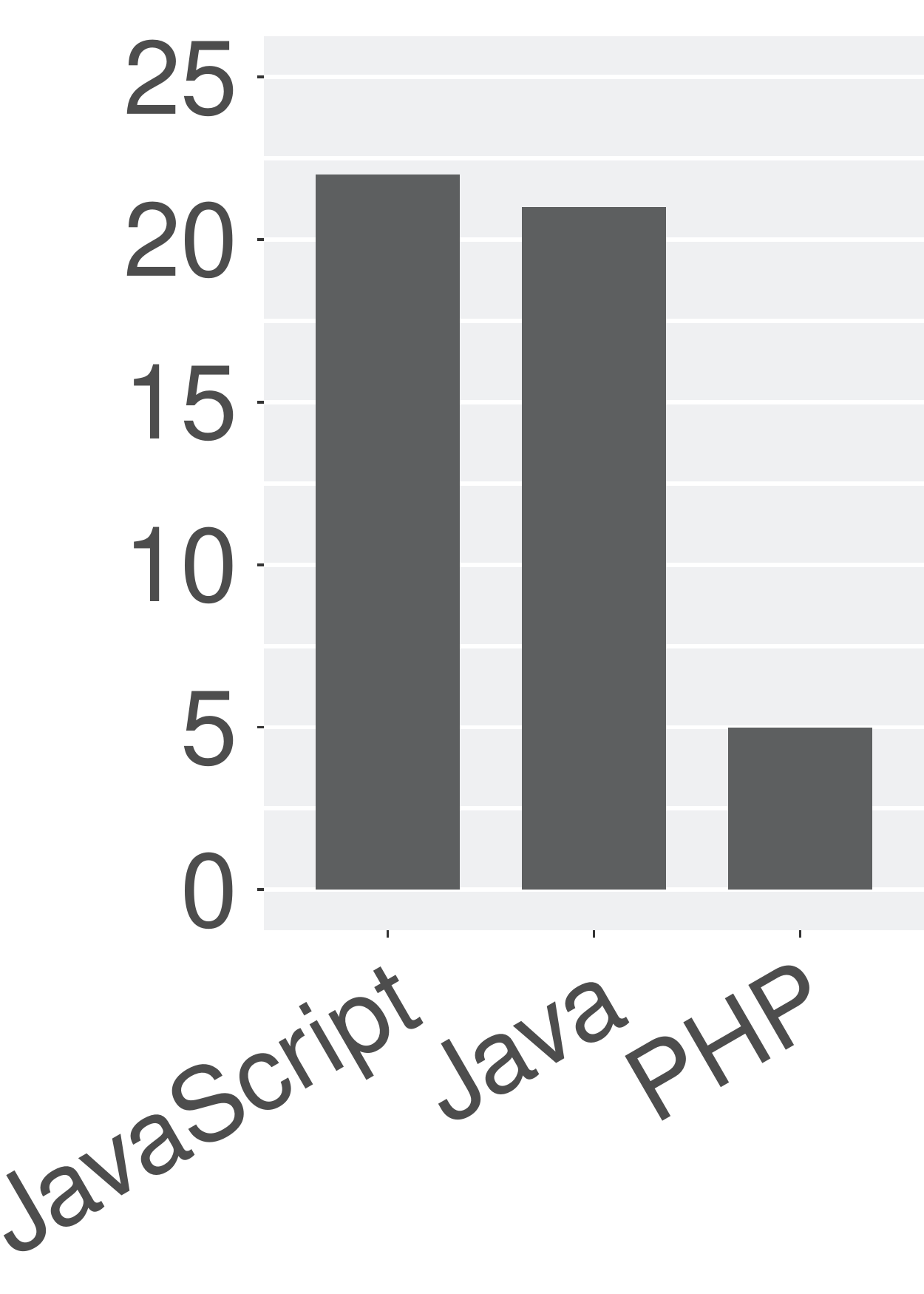}
\label{subfig:language_spain}
}
\subfigure[PL]{
\includegraphics[width=0.17\textwidth]{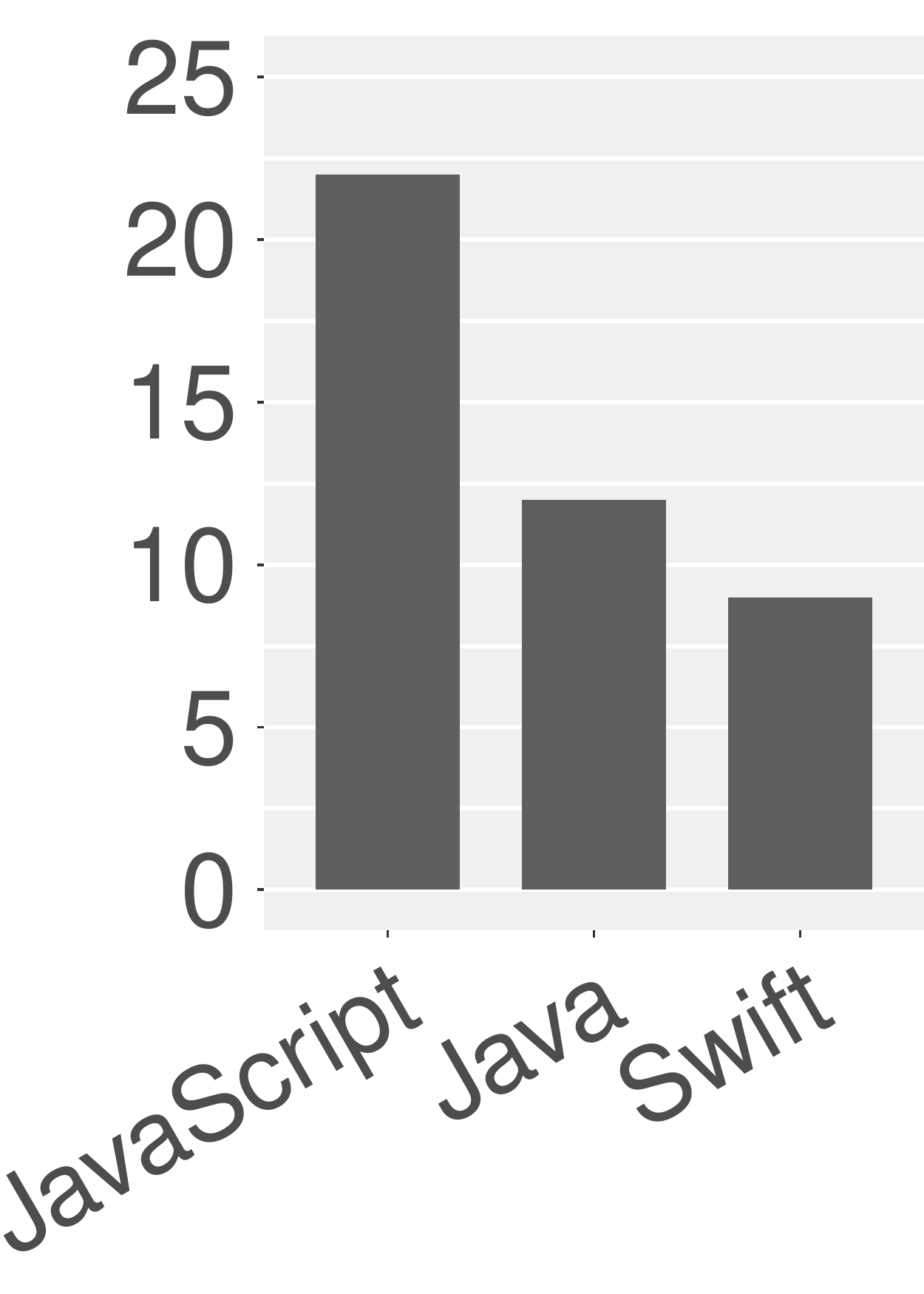}
\label{subfig:language_poland}
}
\subfigure[IT]{
\includegraphics[width=0.17\textwidth]{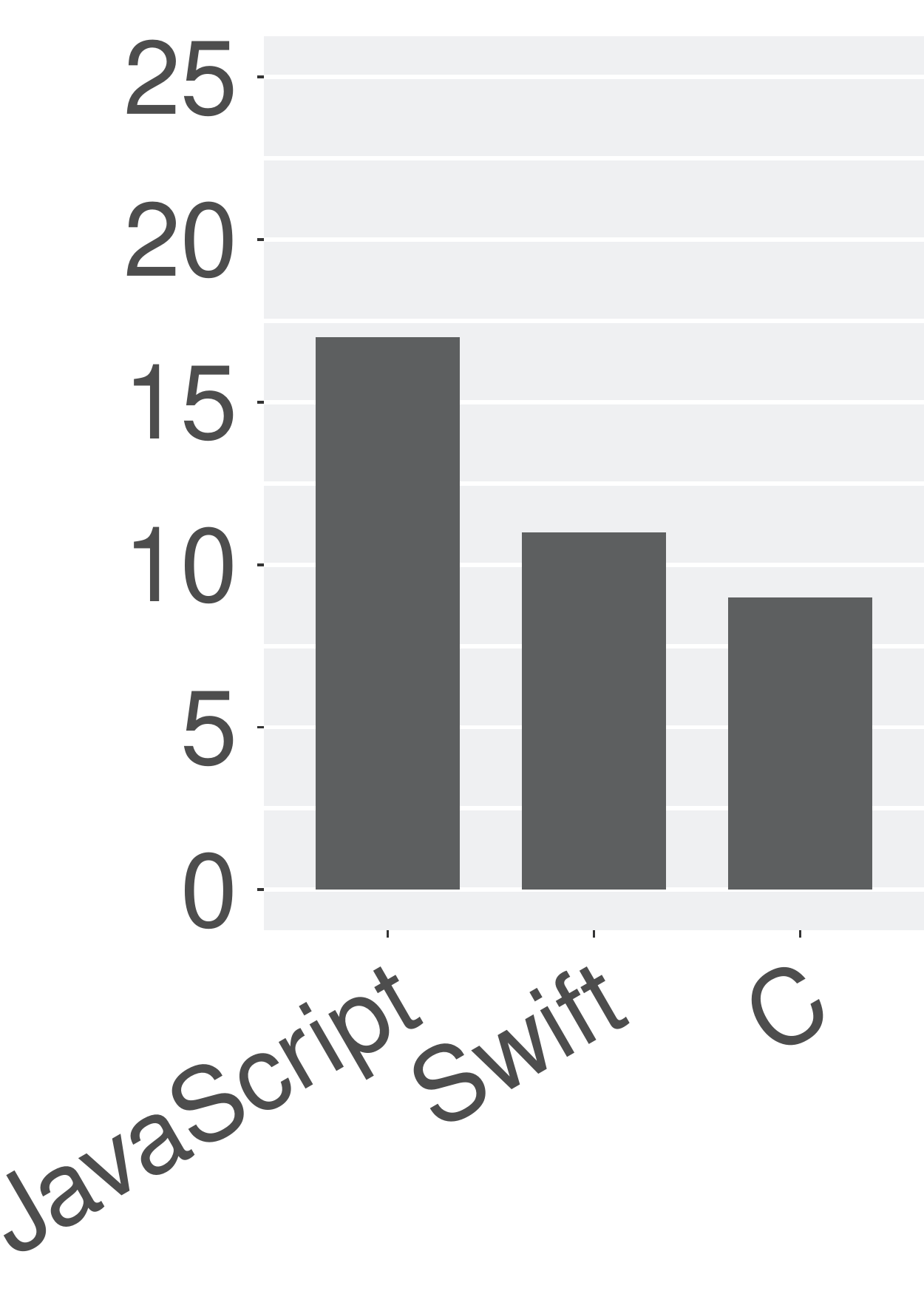}
\label{subfig:language_italy}
}
\subfigure[UA]{
\includegraphics[width=0.17\textwidth]{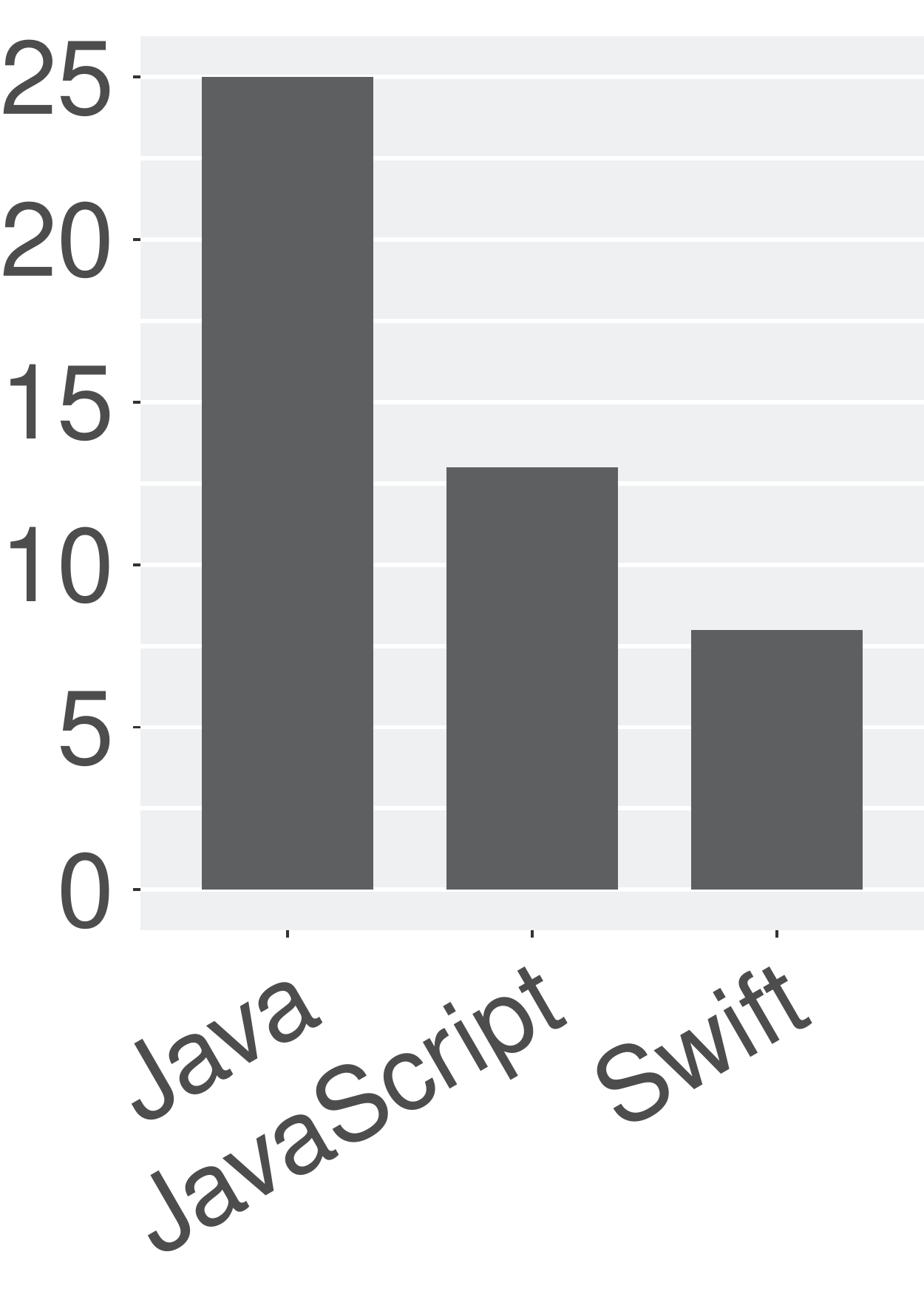}
\label{subfig:language_ukraine}
}
\subfigure[CH]{
\includegraphics[width=0.17\textwidth]{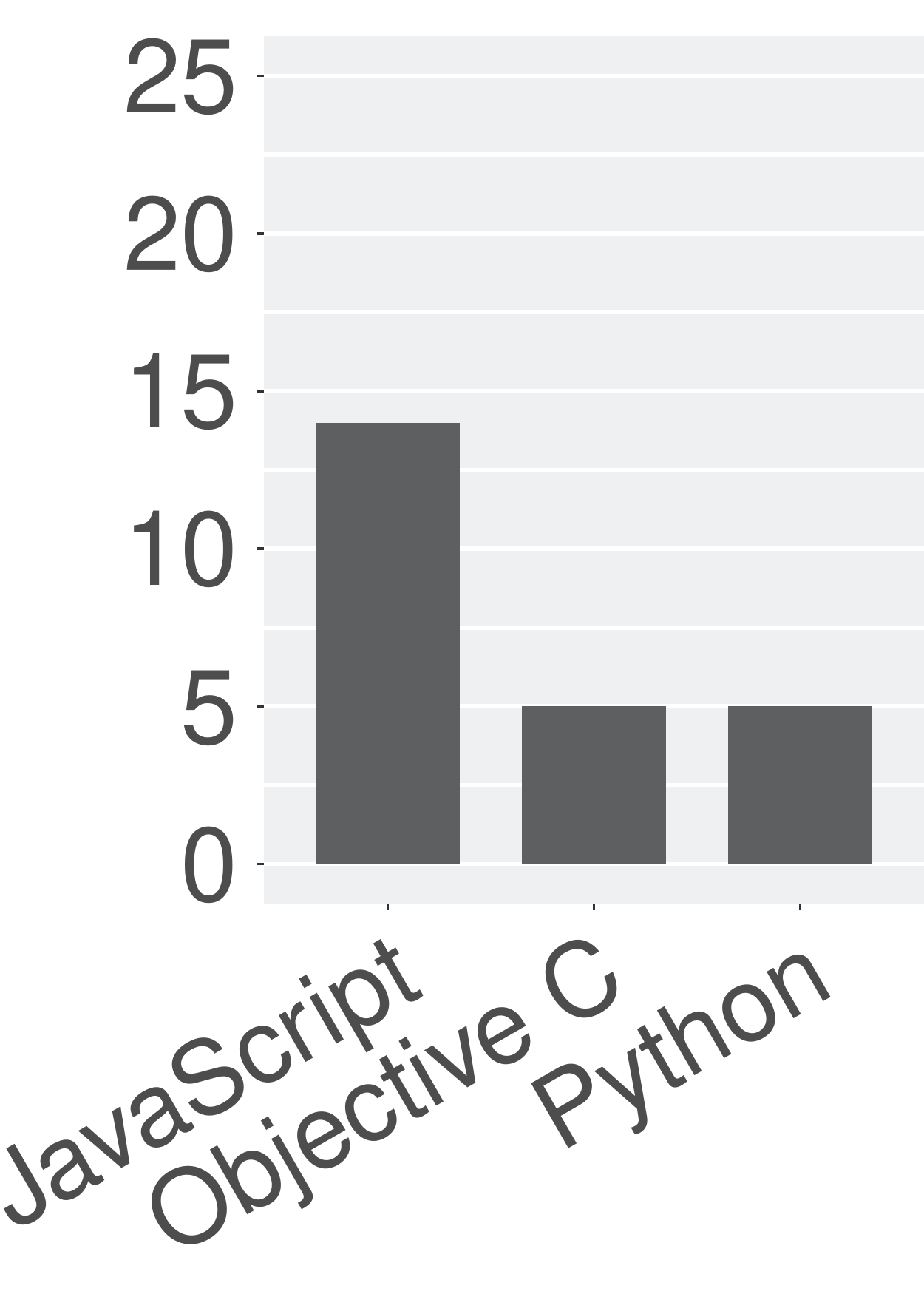}
\label{subfig:language_switzerland}
}
\subfigure[BR]{
\includegraphics[width=0.17\textwidth]{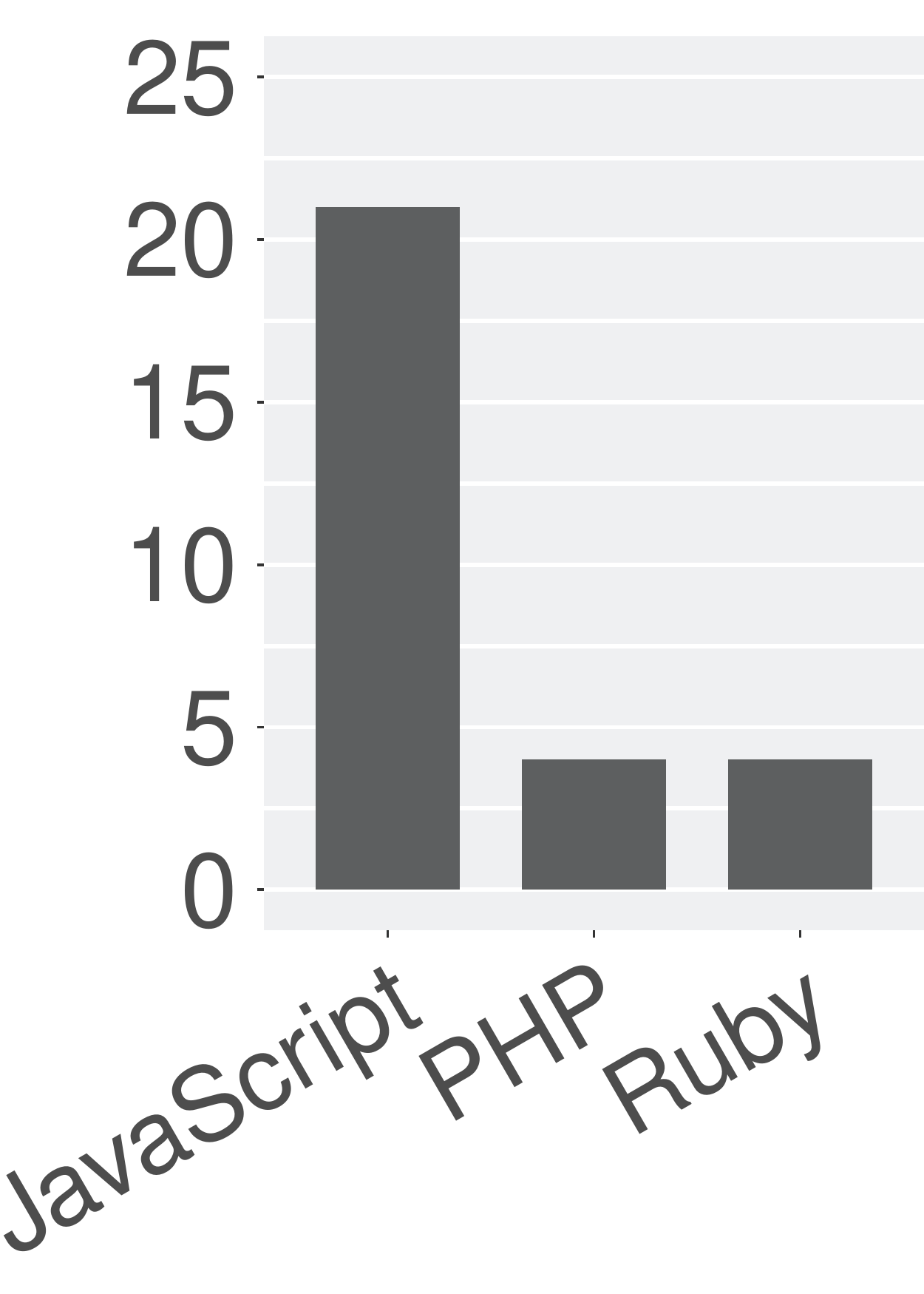}
\label{subfig:language_brazil}
}
\subfigure[AT]{
\includegraphics[width=0.17\textwidth]{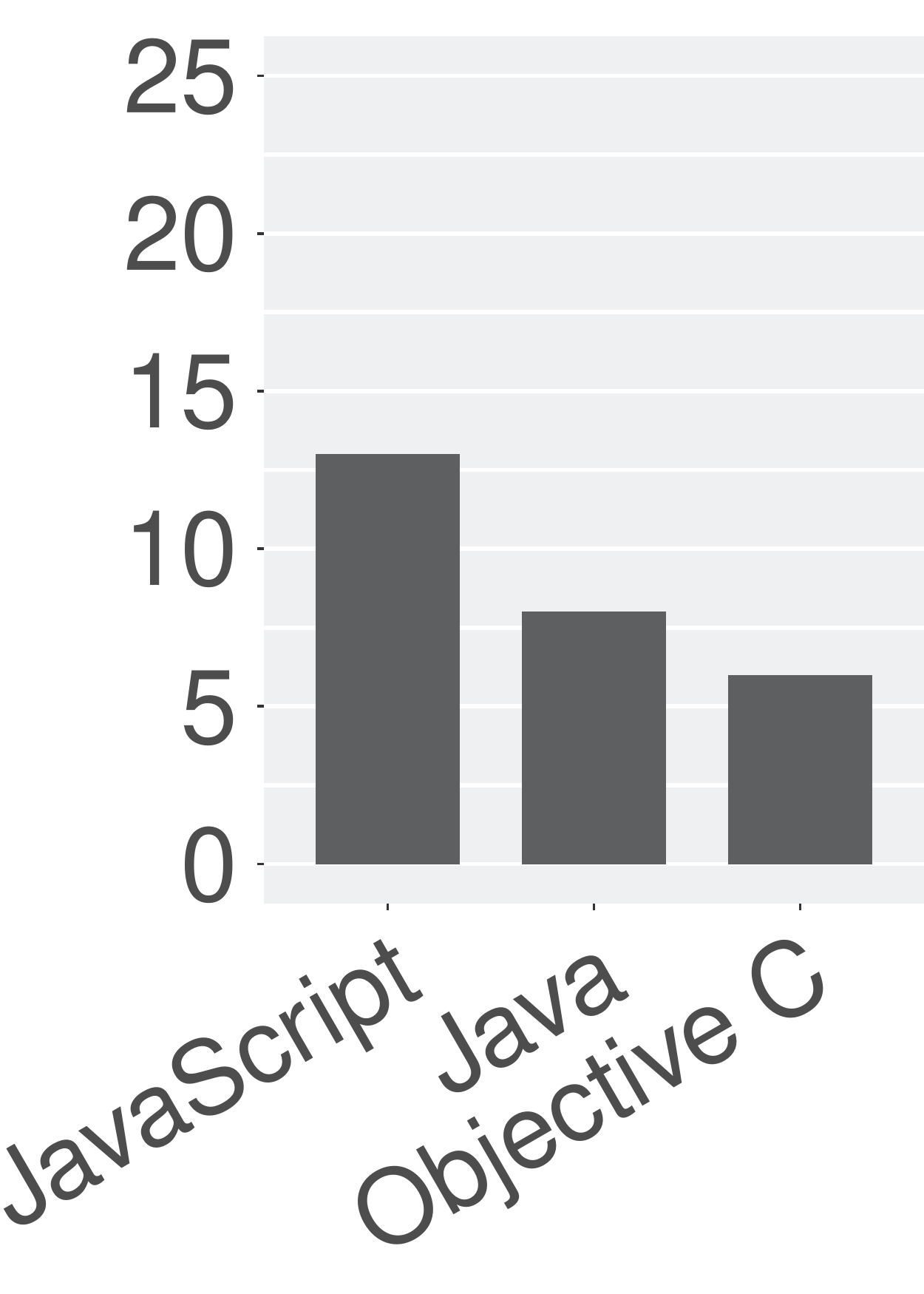}
\label{subfig:language_austria}
}
\subfigure[FI]{
\includegraphics[width=0.17\textwidth]{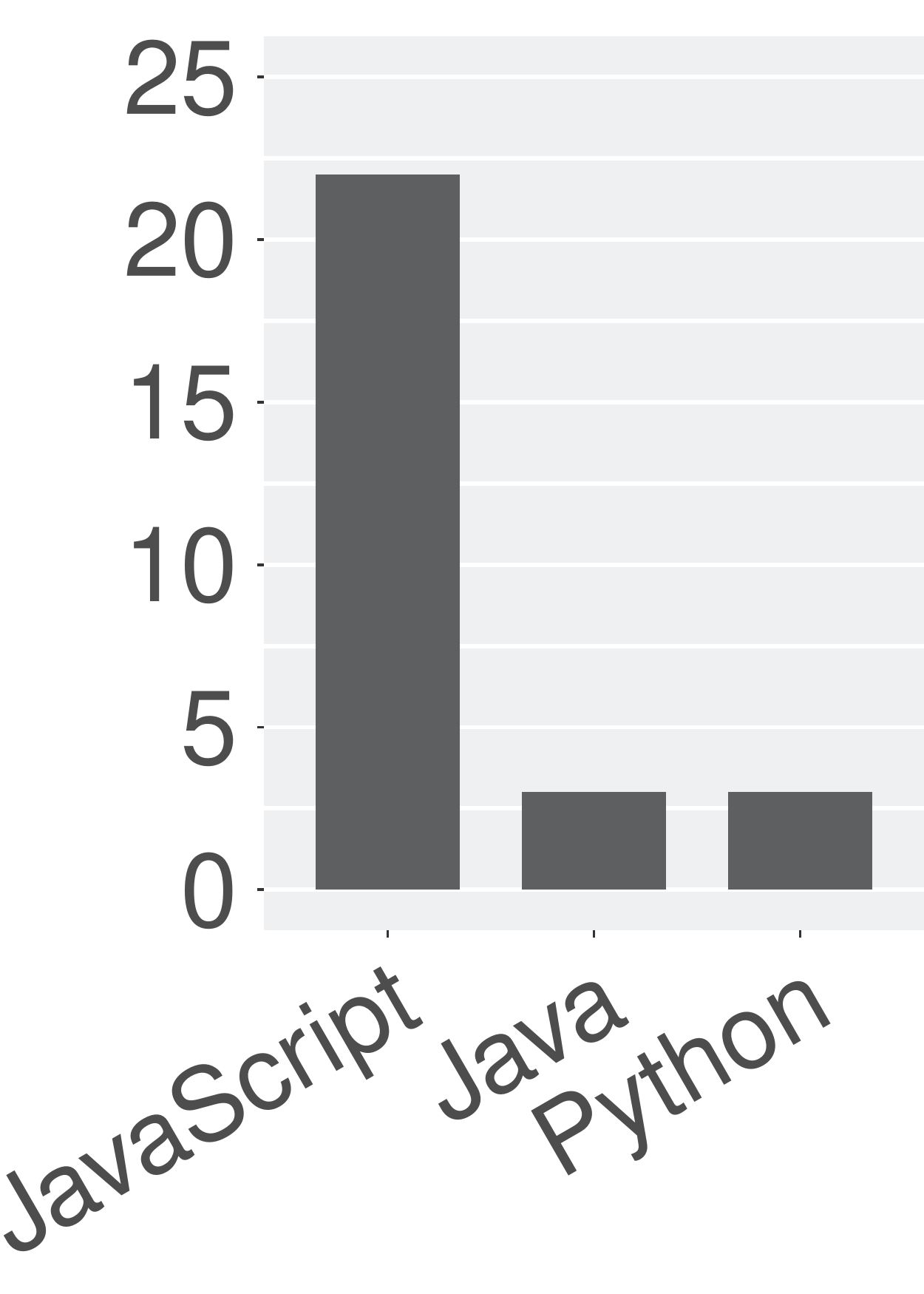}
\label{subfig:language_finland}
}
\subfigure[NO]{
\includegraphics[width=0.17\textwidth]{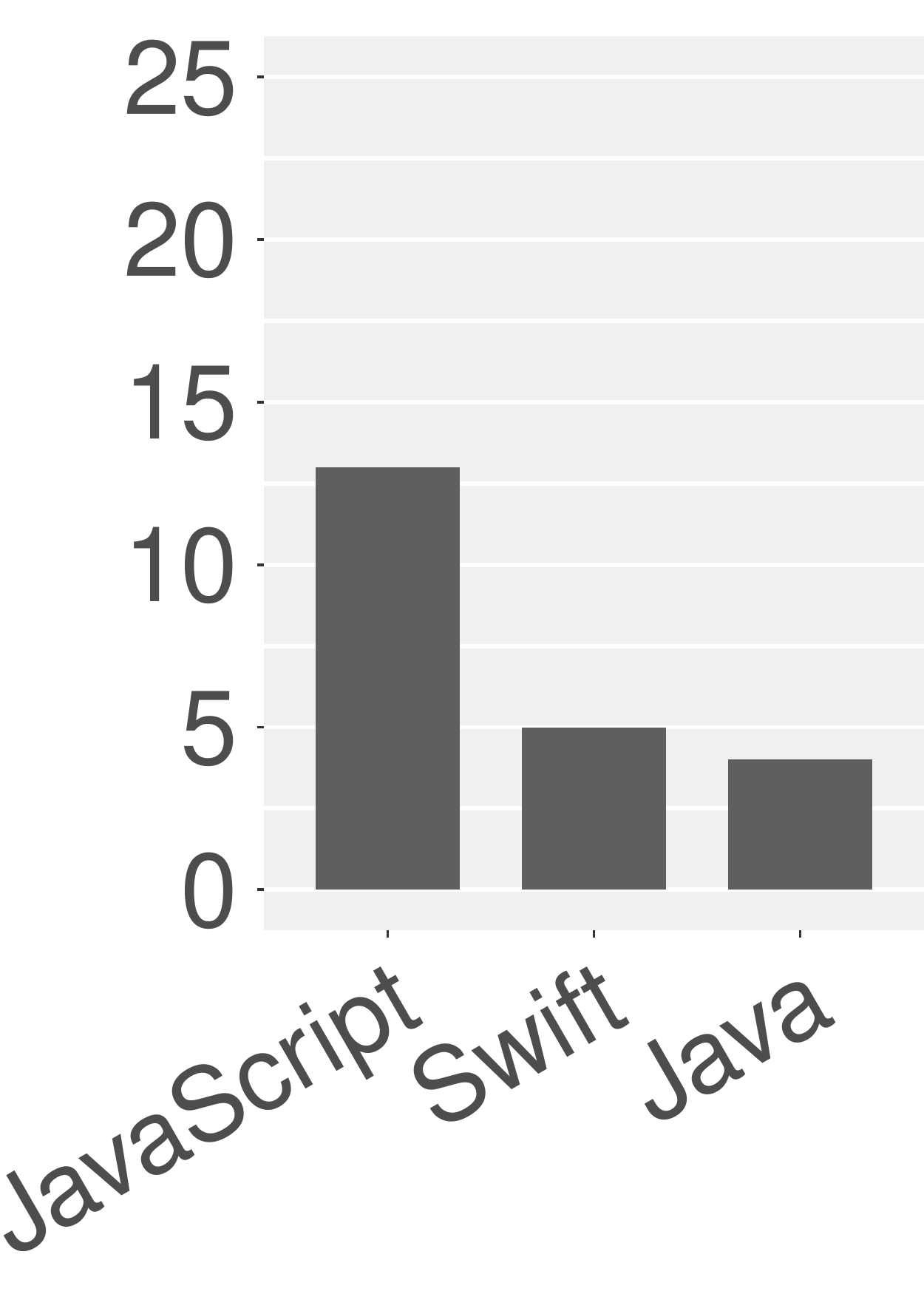}
\label{subfig:language_norway}
}
\caption{Top-3 most popular programming languages (by number of projects)}
\label{fig:language}
\end{figure}

\subsection{International vs Domestic Projects}
\label{sec:international-domestic}

In our work, the owner location is used to define the country of the project.
However, open source projects can attract contributors from other countries. Therefore, we also investigate the percentage of \emph{core developers} who are from a country different from the project's owner. We focus on core developers because it is well-known that open source projects may include several contributors, but only a small number are responsible for the bulk of the implementation and maintenance work~\cite{jergensen2011onion}. 
In this paper, we use a standard heuristic to identify core developers: they are the smallest subset of developers responsible for at least 80\% of the project commits (excluding merges)~\cite{joblin2017classifying}. Moreover, a core developer should have at least 5\% of the projects commits; this is important to avoid including in the core team developers with small contributions, but who are required to reach the 80\% threshold. We defined the 5\% threshold after 
some experiments; for example, the traditional heuristic results in 22 core developers for {\sc spotify/luigi}; however, 18 have less than 5\% of the commits in this system. By contrast, the literature reports that even in complex projects, the core team is no larger than 10-15 developers~\cite{mockus2002two}.  

Using these concepts, we classify the projects of each country in three groups: (a) {\em domestic projects}, when all core developers are from the same country of the project owner; (b) {\em international projects}, when at least one core developer is from a country distinct from the project owner; (c) {\em undefined}, when it is not possible to infer the country of at least one of the core developers (e.g., because the location field is empty or invalid) and the remaining developers are from the same country of the project owner.
Figure~\ref{fig:ini_countries} shows the distribution of projects in the proposed categories, for each country. Domestic projects is the most common category in 19 countries (the only exception is Finland). The Asian countries have the highest percentage of domestic projects, including China (76\%) and Japan (73\%). By contrast, the country with the highest percentage of international projects is Finland (47\%), followed by Canada (41\%) and Brazil (40\%). The percentage of undefined projects is at most 17\%  in all countries, with the exception of Ukraine (32\%). Essentially, our results show that open source is frequently the outcome of a local effort; therefore, measures and policies restricted to a country (e.g., national events) have the potential to promote open source practices in this country.

\begin{figure}[!ht]
\centering
\includegraphics[width=\textwidth]{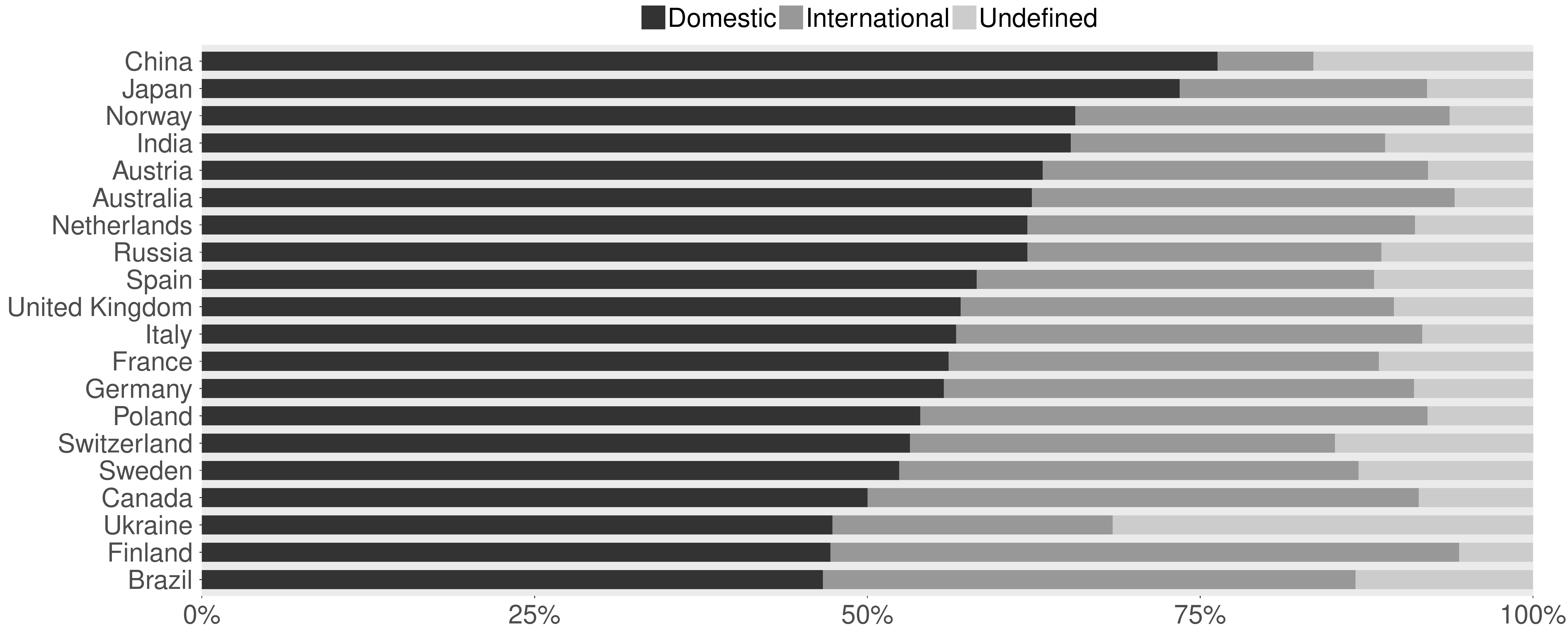}
\caption{International and domestic projects per country}
\label{fig:coredevs}
\end{figure}

\subsection{Maintainability}
\label{sec:bch}

Maintainability is a key concern in modern open source development, particularly in the case of projects maintained by independent developers, without stable financial
support. For this reason, we also analyze how maintainability varies across the projects in each
country. To this purpose, we relied on BetterCodeHub (\url{https://bettercodehub.com}), a web-based tool that assesses
 the maintainability of GitHub projects. BetterCodeHub checks a codebase for compliance with 
 a set of guidelines, covering key aspects of source code quality. 
 For example, the tool
 checks the size and complexity of methods (or similar program elements),
 the presence of duplicated code, the complexity of unit interfaces, the test automation and architecture metrics, among other aspects, implemented for 16 programming languages. The thresholds to check compliance with the proposed guidelines are derived from an ultra-large dataset of software
 projects, totaling 9 billion lines of code, in a benchmark of more than 200 languages.
As its principal result, BetterCodeHub provides a maintainability score for a project. The tool checks for ten guidelines, but in this study we discarded two guidelines that require developer validation of the files that are part of the architectural components in a project. This decision was important to allow scaling our analysis to thousands of systems. Therefore, the scores produced by BetterCodeHub in our setup ranges from 0 to 8; the higher the score, the higher the project's maintainability. Finally, it is  important to highlight that maintainability is one aspect of software quality; although other aspects are also important, such as reliability, usability, portability, etc, they are not part of our study's scope.

We executed BetterCodeHub over 2,648 repositories and received the analysis results on 2,500 repositories. BetterCodeHub did not analyze 148 repositories due to unsupported technologies, oversize, or very long analysis time. Figure~\ref{fig:bettercodehub} shows the distribution of the maintainability scores computed by the tool. The median score is 5 for 14 countries. The six other countries have a median score of 4. The third quartile is 6 for 12 countries; Poland has the highest third quartile measure (score 7). We also found 86 projects with a maximal score (3.4\%), distributed over 17 countries.

\begin{figure}[!ht]
\centering
\includegraphics[width=\textwidth]{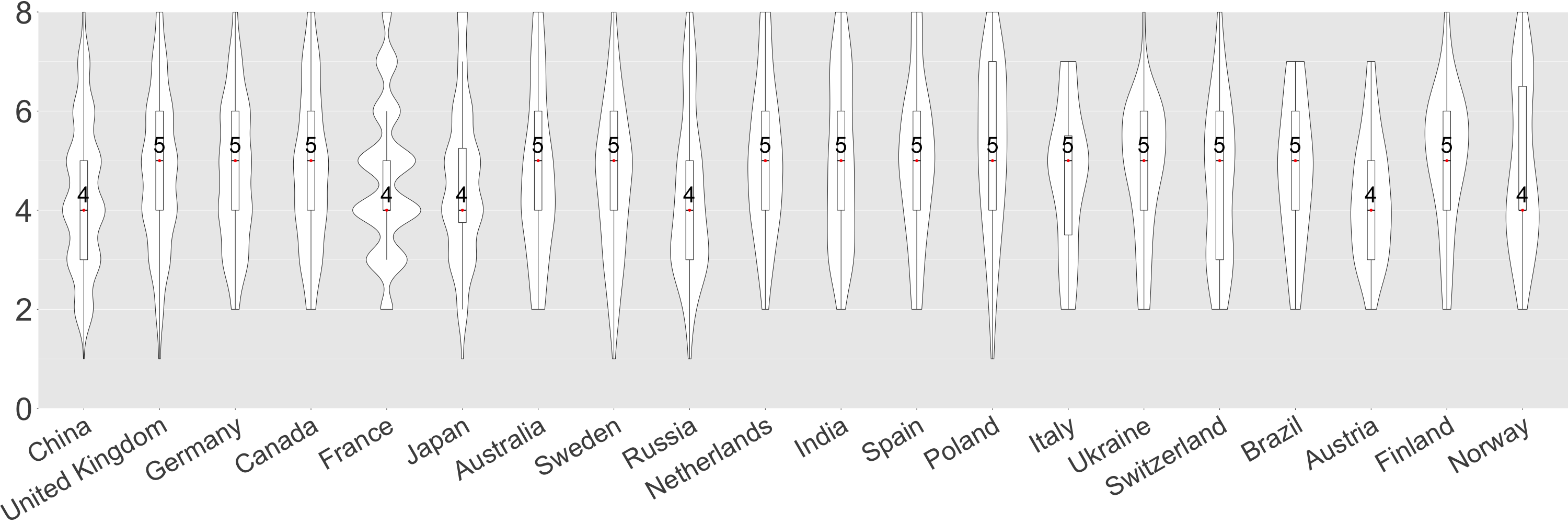}
\caption{Maintainability scores as computed by BetterCodeHub (ranging from 0 to 8)}
\label{fig:bettercodehub}
\end{figure}

Figure~\ref{fig:guidelines} shows the percentage of projects following each of the guidelines checked by BetterCodeHub. The two guidelines with the highest compliance rates are {\em small codebase} (99\%, suggesting that most studied systems have volume less than 20 man-years, which indicates a proper size to ease maintenance) and {\em clean code} (96\%, suggesting the projects have few code smells, which also ease maintenance).
By contrast, the three guidelines with the lowest compliance rates are {\em simple units} (35\%, suggesting that code with complex execution paths, including several branches and loops, is common), {\em automated tests} (26\%, revealing the studied projects  have an insufficient number of tests), and {\em short units} (21\%, showing that long methods or functions are common). 


\begin{figure}[!ht]
\centering
\includegraphics[width=\textwidth]{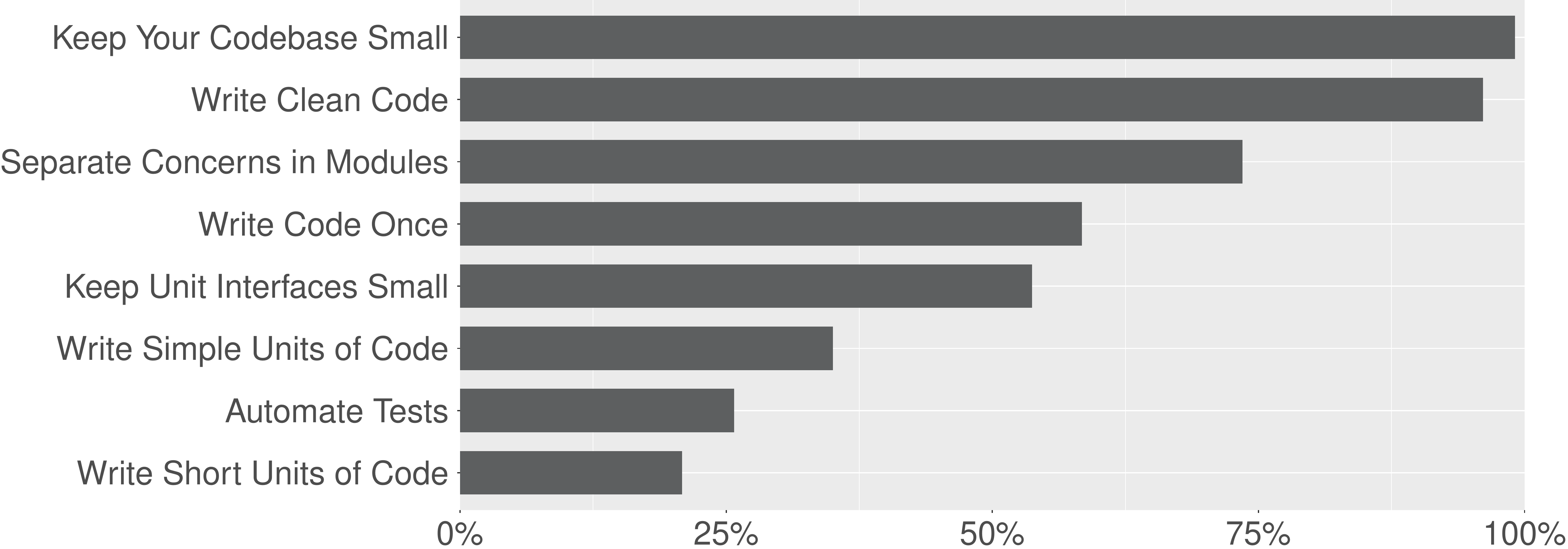}
\caption{Percentage of projects following the BetterCodeHub guidelines}
\label{fig:guidelines}
\end{figure}


Overall, we found that there is no major differences among the studied countries, concerning their final maintainability scores, as computed by BetterCodeHub and presented in Figure~\ref{fig:bettercodehub}. However, this result suggests that there is space for improvements, specially regarding the following guidelines: write simple units, use automated tests, and write short units.



\section{Limitations}

First, the study is restricted to GitHub; although GitHub is the largest platform for developing open source projects, we acknowledge that particular countries might have preferences for other platforms. Another limitation is related to the fact that the location field on GitHub is a free form (i.e.,~it accepts any information). Therefore, our approach for inferring the location cannot be fully automated. The first author had to double check 3,174 locations that our algorithm was not successful to infer. This step was necessary because, even though we could infer the name of well-known cities and states (e.g., Belém is a city in Brazil), acronyms (e.g., BL is an acronym for Belém) and typos (e.g., some developers declare Belemm, with a double m and without accent) are also common. Since we analyzed thousands of projects, such minor issues, at scale, make a fully automated technique unfeasible. Still, our approach can also introduce false positives. For instance, Belém is also a city in Portugal. 

\section{Related Studies}
								
The recent introduction of social coding platforms has drastically changed the way development teams communicate and collaborate. As a side effect of this introduction, open source software is gaining momentum in both practitioners and researchers arenas~\cite{popularity2016,terrell2017gender,ray2014large,dyer2014mining,steinmacher2015systematic,nistor2015c}. On the technical side, researchers are taking advantage of these platforms to answer questions regarding the adoption of programming language features~\cite{dyer2014mining}, how to detect and fix bugs~\cite{nistor2015c}, or how to improve code quality~\cite{ray2014large}. The findings of these studies provide practical and timely implications to several stakeholders, including developers, tool builders, and even programming language designers. Moreover, other studies focused on the social side, such as the characterization of open source communities~\cite{terrell2017gender} and factors that impact developers onboarding and retention~\cite{steinmacher2015systematic} of new contributors in open source communities. Gender bias is also being explored lately~\cite{terrell2017gender}. The popularity and attractiveness of open-source projects have also been a subject of investigation~\cite{popularity2016}. These studies provide some light on the barriers related to diversity and inclusion in open source communities, how can community members mitigate these barriers, and how does open source software contribute to build a more equitable society.

However, to best of our knowledge, there is little evidence on whether a project location (or country) has influence on metrics such as popularity, programming language adoption, or code quality. 
In this paper, we contribute to technical and social axes. 
We provide evidence about the most used programming languages whilst highlighting the role that the contributors play in the studied software projects.

\section{Conclusion}

In this paper we provide an extensive investigation of over 2,648 open source projects developed in 20 countries. We observed that, after removing US projects, the countries with more popular projects are China (754 projects), United Kingdom (336 projects), and Germany (236 projects). We also found a correlation between number of projects and a country's GDP. We found that the most popular programming languages are JavaScript and Java, but other languages, such as Objective C, Swift, and Go, are popular in specific countries.
Moreover, domestic projects, i.e., when all core developers are from the project's country, are the most common category in 19 out of 20 countries. Finally, we showed that there is an space to improve projects' maintainability in all studied countries. Our findings may help governments, entrepreneurs and non-profit organizations to understand open source practices in specific countries, and therefore plan actions to improve them.

\small
\bibliographystyle{unsrt}
\bibliography{sample}

\end{document}